\documentclass[manuscript,screen]{acmart}

\usepackage{subcaption}
\usepackage{multirow}
\usepackage[bottom]{footmisc}
\usepackage{diagbox}
\usepackage{dblfloatfix}
\usepackage[title]{appendix}

\usepackage[font=normalsize,labelfont=bf]{caption}

\AtBeginDocument{%
  \providecommand\BibTeX{{%
    \normalfont B\kern-0.5em{\scshape i\kern-0.25em b}\kern-0.8em\TeX}}}

\setcopyright{acmcopyright}
\copyrightyear{2018}
\acmYear{2018}
\acmDOI{XXXXXXX.XXXXXXX}

\acmConference[Conference acronym 'XX]{Make sure to enter the correct
  conference title from your rights confirmation emai}{June 03--05,
  2018}{Woodstock, NY}
\acmBooktitle{Woodstock '18: ACM Symposium on Neural Gaze Detection,
 June 03--05, 2018, Woodstock, NY} 
\acmPrice{15.00}
\acmISBN{978-1-4503-XXXX-X/18/06}

\begin{document}

\title{Diagnosing Data from ICTs to Provide Focused Assistance in Agricultural Adoptions}


\author{Ashwin Singh}
\affiliation{%
  \institution{International Institute of Information Technology, Hyderabad}
  \country{India}}
\orcid{0000-0002-2596-9649}
\email{ashwin19.iiith@gmail.com}

\author{Mallika Subramanian}
\affiliation{%
  \institution{International Institute of Information Technology, Hyderabad}
  \country{India}}
\email{mallika.subramanian@students.iiit.ac.in}

\author{Anmol Agarwal}
\affiliation{%
  \institution{International Institute of Information Technology, Hyderabad}
  \country{India}}
\email{anmol.agarwal@students.iiit.ac.in}

\author{Pratyush Priyadarshi}
\affiliation{%
  \institution{International Institute of Information Technology, Hyderabad}
  \country{India}}
\email{pratyush.priyadarshi@students.iiit.ac.in}

\author{Shrey Gupta}
\affiliation{%
  \institution{International Institute of Information Technology, Hyderabad}
  \country{India}}
\email{shrey.gupta@students.iiit.ac.in}

\author{Kiran Garimella}
\affiliation{%
  \institution{School of Communication and Information, Rutgers University}
  \country{USA}}
\email{kiran.garimell@rutgers.edu}

\author{Ponnurangam Kumaraguru}
\affiliation{%
  \institution{International Institute of Information Technology, Hyderabad}
  \country{India}}
\email{pk.guru@iiit.ac.in}

\author{Sanjeev Kumar}
\affiliation{%
  \institution{Digital Green}
  \country{India}}
\email{sanjeev@digitalgreen.org}

\author{Ritesh Kumar}
\affiliation{%
  \institution{Digital Green}
  \country{India}}
\email{ritesh@digitalgreen.org}

\author{Lokesh Garg}
\affiliation{%
  \institution{Digital Green}
  \country{India}}
\email{lokesh@digitalgreen.org}

\author{Erica Arya}
\affiliation{%
  \institution{Digital Green}
  \country{India}}
\email{erica@digitalgreen.org}

\renewcommand{\shortauthors}{Anonymous Authors}

\begin{abstract}
In the last two decades, Information and Communication Technologies (ICTs) have played a pivotal role in empowering rural populations in India by making knowledge more accessible. Digital Green is one such ICT that employs a participatory approach with smallholder farmers to produce instructional agricultural videos that encompass content specific to them. With the help of human mediators, they disseminate these videos to farmers using projectors to improve the adoption of agricultural practices. Digital Green’s web-based data tracker (CoCo) stores the attendance and adoption logs of millions of farmers, the videos screened to them and their demographic information. In our work, we leverage this data for a period of ten years between 2010-2020 across five states in India where Digital Green is most active and use it to conduct a holistic evaluation of the ICT. First, we find disparities in the adoption rates of farmers, following which we use statistical tests to identify the different factors that lead to these disparities as well as gender-based inequalities. We find that farmers with higher adoption rates adopt videos of shorter duration and belong to smaller villages. Second, to provide assistance to farmers facing challenges, we model the adoption of practices from a video as a prediction problem and experiment with different model architectures. Our classifier achieves accuracies ranging from 79\% to 90\% across the five states, demonstrating its potential for assisting future ethnographic investigations. Third, we use SHAP values in conjunction with our model for explaining the impact of various network, content and demographic features on adoption. Our research finds that farmers greatly benefit from past adopters of a video from their group and village. We also discover that videos with a low content-specificity benefit some farmers more than others. Next, we highlight the implications of our findings by translating them into recommendations for providing focused assistance, community building, video screening, revisiting participatory approach and mitigating inequalities. Lastly, we conclude with a discussion on how our work can assist future investigations into the lived experiences of farmers.
\end{abstract}




\maketitle

\section{Introduction}
Over 58\% of India’s population relies on agriculture as a primary source of livelihood \cite{ibef}, yet nearly one in every four of an estimated 146 million farmers~\footnote{The estimate is based on the total number of landholdings in the country as per the \href{https://agcensus.nic.in/document/is2016/air_is_16-17_210121-final_220221.pdf}{2015-16 Agriculture Census of India}. While the number of farmers is likely higher, there is no reliable data about the same as per \url{https://thewire.in/agriculture/indian-agricultures-enduring-question-just-how-many-farmers-does-the-country-have}} in India continues to live below the poverty line.~\footnote{\url{https://www.business-standard.com/article/economy-policy/in-6-years-before-covid-19-average-farm-incomes-rose-59-debt-58-121092800118_1.html}} As shown in a multi-layer ethnography of cotton farmers, one of the key reasons behind their financial turmoil is the lack of reliable knowledge about agricultural practices \cite{stone2007agricultural}. Therefore, bridging this knowledge gap can help prevent poor decision making and helping improve their financial standing. Some of the earliest ICT interventions that attempted to tackle this problem included agriculture-specific television programs and radio broadcasts. However, these modes lacked the information specificity sought out by farmers. To overcome this limitation, Digital Green, in 2008, introduced (i) a participatory process for the production of instructional video content that allows for the representation of local communities and (ii) the use of human-mediated instruction in the dissemination of video and training process \cite{Gandhi}. While the participatory process allows for Digital Green to tailor their content for the local communities, human mediators play a key role in ensuring active engagement with that content. Their initial study in 2007 involved a four-month trial across 16 villages in India, which saw an increase in the adoption of specific agriculture practices by a factor of six-seven times over traditional modes of television programs and radio broadcasts \cite{Gandhi}. 

As of 2021, Digital Green has scaled across three continents and twelve countries, reaching 2.3M rural households globally.~\footnote{\url{https://www.digitalgreen.org/global-impact/}} Recent years have also witnessed a rapid increase in technology ownership across rural areas of India, with over 17.5M farmers registered with the Electronic National Agricultural Market (e-NAM).~\footnote{\url{https://enam.gov.in/web/dashboard/stakeholder-data}} To tap into this proliferation of digital technologies, Digital Green has introduced several extensions, a notable one being an app ‘Videokheti’ \cite{vkheti1, vkheti2}, that allows farmers to rewatch the instructional videos on agricultural practices. To improve the accessibility to their content, Digital Green has also established its presence on YouTube with over 287K subscribers ~\footnote{\url{https://www.youtube.com/user/digitalgreenorg/}} and curated a digital library of videos on its website.~\footnote{\url{https://solutions.digitalgreen.org/videos/library}} As Digital Green continues to scale its presence in a sustainable manner, it presents the ICTD community with opportunities to evaluate its impact from a computational perspective. In our work, we examine the data collected by Digital Green’s web infrastructure (CoCo) over a period of ten years from 2010-2020; CoCo captures the data related to their key processes, including video production, dissemination and the adoption of agricultural practices.~\footnote{While Digital Green’s videos also include topics such as health, livestock, social issues and financial management, we limit our scope to only agricultural practices since the adoption mechanisms for other types of practices can be inherently different.} We perform a holistic diagnosis of Digital Green using CoCo, looking at how various social, temporal and content features influence the adoption of best practices mentioned in the video. In particular, we ask the following research questions:

\begin{itemize}
    \item What are the different factors that distinguish farmers who adopt more than others?
    \item How important are these factors, and how does their importance vary for farmers across different Indian states?
    \item How can we identify farmers who face challenges in adopting farming practices to provide assistance to them?
\end{itemize}

The rest of the paper is organised as follows. Section~\ref{related_work} discusses how our work is situated in relevant literature and overcomes some of its limitations in using large scale social data to understand different factors that influence the adoption of diverse agricultural practices. Section~\ref{ecosystem_dataset} elaborates on the human infrastructure of Digital Green and the various components involved in its functioning. It provides a detailed overview of the data captured that we use in this paper. In Section~\ref{modelling_components}, we model three different components of the Digital Green ecosystem using the data from CoCo --- (i) attendance and adoption behaviours of farmers, (ii) content from the instructional agricultural videos, and (iii) demographic information of farmers. Next, we make use  of these features in Section~\ref{differential_factors} to understand why some farmers adopt more than others. We leverage aggregate statistics from demographic features to diagnose the ICT for potential gender-based inequalities. Following the diagnosis, we model the adoption of agricultural practices from a video as a prediction problem in Section~\ref{predicting_adoptions}, to identify farmers who might face challenges in adoption.
Then, we outline the implications of our findings (Section~\ref{implications}) and reflect on our positions as researchers (Section~\ref{positionality}). Finally, we conclude with a discussion (Section~\ref{discussion}) on how our diagnosis can assist Digital Green in conducting ethnographic research to generate experiential considerations in a supplemental way. This would also help improve their understanding of what interventions work well, for who they work well and the potential pitfalls in implementing them within their ICT.

\section{Related Work} \label{related_work}

Our work is broadly situated in the domain of ICTs for rural development. More specifically, it focuses on the use of ICTs to disseminate knowledge about agricultural practices and data-driven methods for their evaluation. Therefore, it bridges the literature concerning the generation of big data and the adoption of agricultural practices. In the next two subsections, we elaborate on the same in detail.

\subsection{ICTs 4 Rural Development}

In the last two decades, ICTs have played a fundamental role in rural development and the transfer of vital information to empower smallholder farmers. Ranging from topics such as sowing and crop protection to improving soil fertility, knowledge transferred through ICTs has enabled farmers to make more informed decisions and improve their agricultural productivity.~\footnote{\url{https://www.cropin.com/ict-in-modern-agriculture/}} Over the years, this knowledge has been communicated via increasingly diverse modes, including kiosks \cite{ishakti}, web portals~\footnote{url{https://www.ikisan.com/}} and smartphone applications.~\footnote{url{https://www.tatatrusts.org/our-work/livelihood/agriculture-practices/mkrishi}}$^{,}$\footnote{url{https://www.iffcokisan.com/}} Dissemination of videos through these modes has positively impacted the knowledge sharing and communication process among farmers in recent years \cite{4937421, rose2021videos, bentlay, farmingvlogs}. However, the extent of their impact has been largely dependent on how they facilitate social learning \cite{karubanga2017farmer}. For instance, in a conventional top-down transfer of technology, knowledge becomes equivalent to a commodity delivered to farmers, thereby restricting their control over its management \cite{cuellar2011can}. This limits not only their creativity in farming decisions but also makes them more reliant on a predefined set of instructions that are not necessarily curated for them \cite{pagliarino2020participatory, chambers1994farmer}. To overcome this limitation, ICTs have seen a rise in the use of participatory design for the production of video content. First, it allows for the curation of knowledge according to a local community’s requirements \cite{Gandhi, barakabitze2017use}. Second, content sharing through videos created by farmers for farmers has been proved to be far more impactful in facilitating learning of agricultural practices and improving their adoption \cite{zossou2012participatory}. As an ICT, Digital Green not only leverages both these advantages but also amplifies them by tapping into its human infrastructure. Digital Green’s idea of involving human mediators to disseminate relevant information was later recognised as ‘\textit{infomediaries}’ in the ICT4D 2.0 Manifesto \cite{ict4d2}. Similar strategies have been observed across health workers in rural India, where ASHAs (Accredited Social Health Activists) are actively involved in engaging high-status infomediaries in the process of video production \cite{ASHA}. Further, the mediators involved in the dissemination of Digital Green’s videos, along with the actors featuring in them are often farmers from the local community. This enables homophily \cite{Gandhi}, thereby facilitating more engagement with the content. Localization of video content disseminated to farmers in terms of language and dialect has also shown improvement in learning and assimilation \cite{educationalanimated, videomediatedrural}. Success of the participatory approach is noticeable in literature beyond ICTs as well. Employing human-centred designs and adopting involved methodologies at the grass-root level to help educate communities has amplified the performance of frameworks such as Decision Support Systems (DSS) and Knowledge Management Models (KMM) \cite{knoche2010missing, vangala2015icts, lynch2004user}. Despite the success rates of the participatory approach, there continues to remain an inherent creative tension between its adoption and its use to advance public policy goals in terms of sustainable agriculture development. There exists a trade-off between leveraging the benefits of the “bottom-up” participatory approach in utilizing ICTs that facilitates significant progress of individuals versus drawing policies for holistic sustainable development \cite{bruges2008participatory}. Our work acknowledges these trade-offs in the implications of our findings, suggesting how quantitative findings can guide qualitative investigations to generate considerations for their implementation and eventually improve the adoption of practices by farmers.

\subsection{Big Data and Adoption in Agriculture}

Understanding the dynamics of components that influence adoption of agricultural practices can greatly benefit farming communities \cite{mittalagricultural_informationnetworks, adoption_agriculture_innovations, farmer_adoption_review_synthesis}. Broadly, these components can be associated with environmental factors, institutional structures, government influence and the information flow dynamics within a community \cite{categorisation_factors_adoptions}. Across each of these, there is potential for big data production in the agriculture industry. However, as highlighted by Kamilaris et al. \cite{big_data_analytics_review}, the current sources of big data in agriculture are limited to remote and proximal sensing tools, historical records of food and climate data, static databases of geospatial data, surveys conducted by the government and web-based accounts of farmers’ decision-making. First, while some of these sources enable studying macro-level socio-economic and policy indicators at scale \cite{land_use_socioeconomic, socioeconomic_determinants_grass_kenya, socioeconomic_factors_india}, they lack consideration for the social dynamics and ground-level interactions that unfold within farming communities. Second, proximal sensing sources enable IoT and cloud-based innovations that support farmers in agriculture \cite{big_data_analytics_review, OrganizedDA_pride_model} but are dependent on technology ownership and thereby susceptible to deepening the digital divide \cite{promises_pitfalls_ict_agriculture, social_inequalities_india}. Third, there is a lack of big data when it comes to capturing the social dynamics of farmers; one can model them as social networks to study knowledge sharing. While previous works \cite{farmers_social_networks_myanmar, social_networks_ghana, measuring_social_networks} look at various ways of information diffusion in such networks, the scale of their evaluation is limited to small and localised farmer populations sizing upto 500. Lastly, when it comes to predicting and analysing trends in the adoption of farming practices, past works have been limited to specific practices \cite{promoting_adoption, forecasting_irrigation}, highlighting the scope for studying how factors impact a multitude of farming practices. Digital Green’s web infrastructure (CoCo) helps overcome these limitations; it captures the attendance and adoption logs of millions of farmers across the globe for videos containing various types of farming practices. Thus, our work utilizes this large scale data to conduct a holistic evaluation of the ICT.

\section{The Digital Green Ecosystem and Dataset Description} \label{ecosystem_dataset}

Digital Green's (DG) ecosystem consists of various actors and components, which are described in Table~\ref{tab:terminology_actors_attributes}. It starts with the participatory production of video content, where content producers (scientists, NGO experts, field staff and progressive farmers) involve the local farmers in creating instructional videos tailored for the community \cite{Gandhi}. Mediators with varying levels of expertise (frontline workers and extension officers) conduct screenings to disseminate these videos to groups of farmers from the local community. Mediators are also supported by the partners who are employees from the government or NGOs such as Bharatiya Agro Industries Foundation~\footnote{https://baif.org.in/} and Samaj Pragati Sahayog.~\footnote{http://www.samajpragatisahayog.org/} These partners enable feedback and audit mechanisms for clusters of villages. Approximately two weeks after a screening is conducted, the staff associated with DG go on-site to survey farmers along with the mediator and associated partner to verify the adoption of practices disseminated in the videos. A video consists of three to five key recall points corresponding to each practice, which are either verified physically or through a knowledge recall method by the surveyor. These surveyors report back to centres and the data entry operators input this adoption data into CoCo. 

\begin{table}[!h]
  \caption{\textbf{Terminology for actors and components as per Digital Green.}}
  \label{tab:terminology_actors_attributes}
  \begin{tabular}{lp{0.7\linewidth}}
    \toprule
    Actor/Component & Description\\
    \midrule
    Farmer & A person pursuing farming as a member of Digital Green.\\
    Group & Self help groups of farmers formed by the government.\\
    Partner & An NGO or government organisation associated with the activities of DG.\\
    Mediator & A frontline worker or officer who disseminates the videos to farmers.\\
    Video & A video containing practices relevant to the farmers.\\
    Screening & Screening of relevant video(s) via a projector to target groups of farmers.\\
    Adoption & Verified instances of farmers implementing or learning practices from the video.\\
  \bottomrule
\end{tabular}
\end{table}

\begin{table}[!ht]
  \caption{\textbf{Dataset description and statistics, showing the scale of the CoCo web infrastructure. (*) Numbers only represent the unique videos screened and adopted. A video can be screened and adopted multiple times and across different states as shown in the Venn diagram (see Figure~\ref{fig:1}).}}
  \label{tab:dataset_description_stats}
  \resizebox{\textwidth}{!}{%
  \begin{tabular}{cccccccccc}
    \toprule
    State & Districts & Blocks & Villages & Groups & Farmers & Mediators & Screenings & Videos  Screened & Videos Adopted\\
    \midrule
    Bihar & 38 & 243 & 4,908 & 46,621 & 534,507 & 5,021 & 232,994 & 369 & 343 \\
    Andhra Pradesh & 17 & 288 & 2,951 & 18,274 & 221,052 & 2,409 & 90,163 & 322 & 297\\
    Odisha & 7 & 32 & 1,262 & 6,773 & 107,665 &553 & 78,461 & 461 & 380\\
    Madhya Pradesh & 17 & 72 & 1,295 & 4,480 & 67,478 & 801& 57,739  & 742 & 620\\
    Karnataka & 15 & 35 & 780 & 4,074 & 50,923 &624 & 32,719 & 340 & 266\\
    \hline
    Total & 94 & 670 & 11,196 & 80,222 & 981,625 & 9,408 & 259,082 & 2,208* & 1,896*\\
    \bottomrule
  \end{tabular}}
\end{table}

CoCo contains the data for agricultural screenings and adoptions of videos across a period of ten years between 2010 and 2020. For our work, we focus on only the top five of twelve states in India where DG is most active in terms of the number of screenings conducted -- Bihar, Andhra Pradesh, Odisha, Madhya Pradesh and Karnataka. These states are divided into districts which are further divided into blocks that are constituted by villages at the lowest level. A comprehensive view of the descriptive statistics such as the number of videos, screenings, adoptions, farmers and geographic distribution for these five states is presented in Table~\ref{tab:dataset_description_stats}. 

\begin{figure}[tp!]
\centering
\begin{minipage}{.49\textwidth}
  \begin{minipage}{0.49\linewidth}
        \centering
        \includegraphics[width=0.9\linewidth]{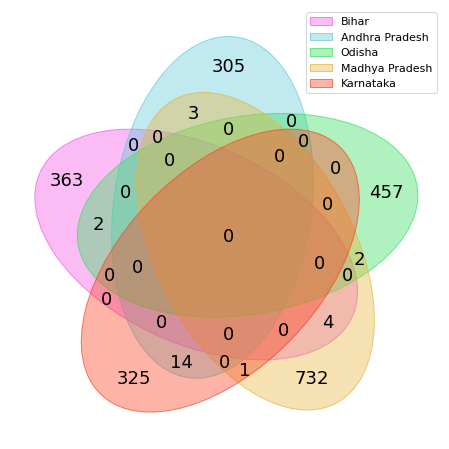}
        \subcaption[first subcaption]{Videos Screened}\label{fig:1a}
    \end{minipage} \hfill
    \begin{minipage}{0.49\linewidth}
        \centering
        \includegraphics[width=0.9\linewidth]{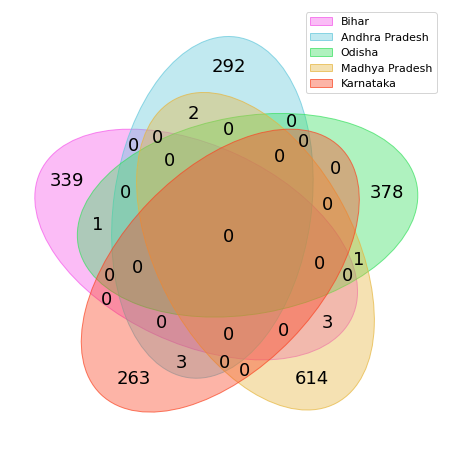}
        \subcaption[second subcaption]{Videos Adopted}\label{fig:1b}
    \end{minipage}
    \caption{
    \textbf{Venn diagram of (a) videos screened and (b) videos adopted across the five states. Of 461 videos in Odisha, only two were screened in Bihar and Madhya Pradesh each, and only one was adopted in each state, demonstrating high content specificity and DG’s community-based approach.}}
  \label{fig:1}
\end{minipage}
\hfill
\begin{minipage}{.45\textwidth}
  \centering
  \includegraphics[width=.8\linewidth]{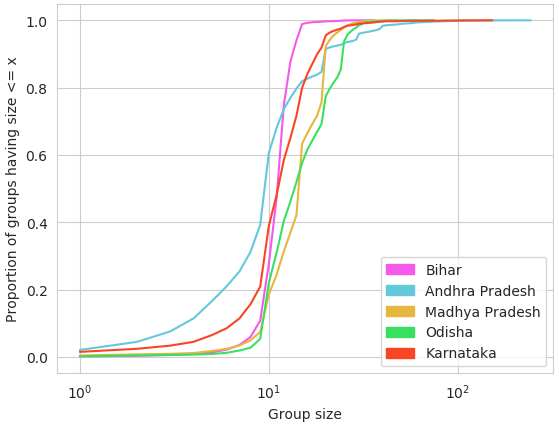}
  \captionof{figure}{\textbf{CDF plot for group sizes ($\mu$=12.24, $\sigma$=6.05). Eighty one percent of the groups comprise 10-30 farmers to ensure a healthy mediator to farmer ratio.}}
  \label{fig:2}
\end{minipage}%
\end{figure}

One of the salient features of DG is its participatory approach to video production i.e., their content is highly tailored to the local communities involved in its production. Figure~\ref{fig:1} shows a Venn diagram of the unique videos screened and adopted across the five states. We observe that most of these videos are specific to the states and the adoption of videos is particularly low at their intersections, highlighting DG's local community-based approach. To quantify it further, we look at this specificity of videos at the village level and find that out of all pairs of villages in each state, the percentage of village pairs adopting at least one common video is very low --- Bihar (9.6\%), Andhra Pradesh (26.4\%), Odisha (11.3\%), Madhya Pradesh (12.1\%) and Karnataka (8.3\%). This demonstrates the high specificity of DG’s content across the five states, even at the lowest geographical level. To disseminate this specific content, farmers are divided into self help groups as beneficiaries of government schemes. DG's videos are targetted towards these groups, therefore, farmers belonging to the groups attend screenings of the same videos together. There are a total of 80,222 (Column 5, Table~\ref{tab:dataset_description_stats}) such groups across the five states with varying sizes ($\mu$ = 12.24, median = 12, $\sigma$ = 6.05). The CDF plot (Figure~\ref{fig:2}) of the group sizes for each of the five states depicts that a large percentage of the farmer groups (81.74\%) comprise 10-30 farmers. 


\begin{figure}[tp!]
\centering
\begin{minipage}{.45\textwidth}
  \centering
  \includegraphics[width=1\linewidth]{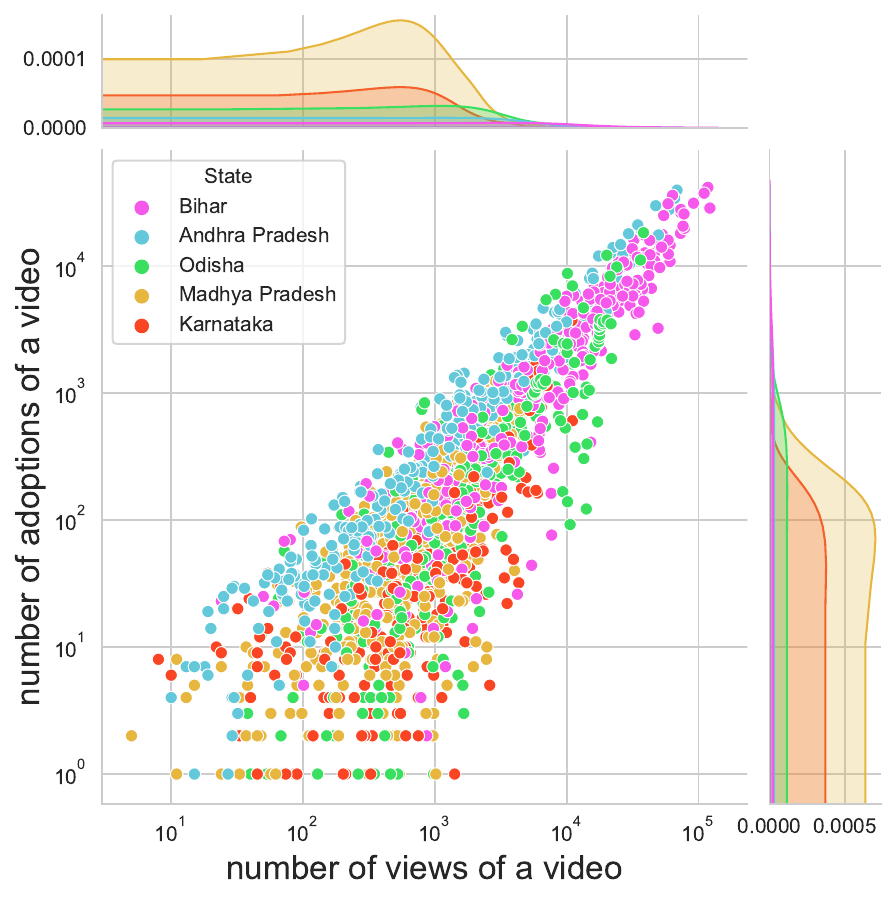}
  \captionof{figure}{\textbf{Joint plot capturing distribution and log-linear trend of adoptions with number of views of a video. Fewer videos are widely adopted as visible from the distribution and funnel shape of the plot.}}
  \label{fig:3}
\end{minipage}
\hfill
\begin{minipage}{.45\textwidth}
  \centering
    \includegraphics[width=1\linewidth]{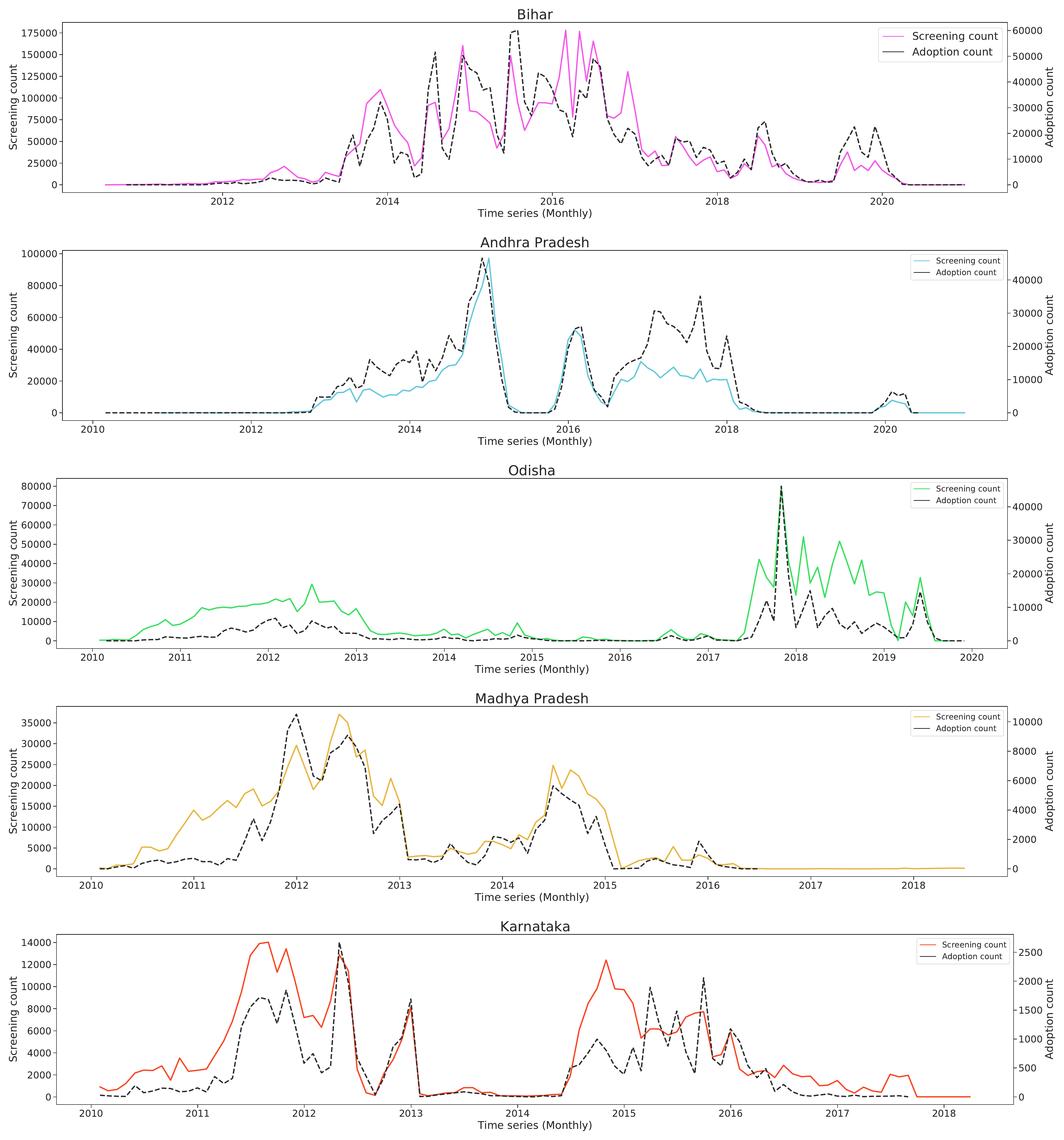}
    \caption{\textbf{\textbf{Timeseries trends for screenings and adoptions across the five states (top to bottom): Bihar, Andhra Pradesh, Odisha, Madhya Pradesh and Karnataka. The spikes in adoptions of farmers almost coincide with the spikes in the screenings of videos, depicting how the adoption behaviour of farmers loosely mimics the trends in screening by DG.}}}
    \label{fig:4}
\end{minipage}%
\end{figure}

We examine the videos closely in Figure~\ref{fig:3} and find that the number of adoptions for a video follows a log-linear trend with its number of views because only a fraction of the viewers adopt the video. The videos in the plot funnel towards the end, denoting that there are only a few videos that are widely adopted. We investigate the video-screening and adoption behaviours further by looking at temporal patterns. In Figure~\ref{fig:4}, we plot the time-series trends for both behaviours. We notice that the spikes in adoptions of farmers almost coincide with the spikes in the screenings of videos across a period of ten years. \\

\noindent\textbf{Ethical Considerations for Data Use}: The dataset made available by Digital Green was collected over the years with the prior consent of the farmers and mediators involved in its functioning. For the purpose of our research, we do not use any personally identifiable information about the actors involved. While the dataset provides us with information about the gender of farmers and mediators, we only make use of it as an aggregated statistic to diagnose the ICT for potential inequalities. Being a sensitive attribute, we avoid using it as a feature in our prediction model. We also take special care while interpreting the quantitative findings from our research and explicitly state the reliance on qualitative fieldwork in implementing them to prevent any discriminatory outcomes.



\section{Modelling Components of the Digital Green Ecosystem} \label{modelling_components}

In this section, we motivate and model relevant features corresponding to the three components present in CoCo -- (i) attendance and adoption behaviours of farmers, (ii) content details of the videos screened to farmers, and (iii) demographic features of farmers and mediators.

\subsection{Attendance and Adoption Behaviours} \label{attendance_adoption_features}

We model the attendance and adoption behaviours of farmers using two temporal networks -- (i) $G_{1} = (F,E)$ where $F$ denotes the set of nodes (farmers) and $E$ denotes the set of edges where an edge $(f,g,w,d) \in E$ represents two farmers $f,g \in F$ who have co-attended $w$ screenings of agricultural videos prior to date $d$, and (ii) $G_{2} = (F,E)$ where an edge $(f,g,w,d) \in E$ represents two farmers $f,g \in F$ who have co-adopted agricultural practices from $w$ videos before date $d$. Due to the high specificity of content in CoCo as seen in Section~\ref{ecosystem_dataset}, we restrict the set of nodes $F$ to farmers belonging to the same village i.e., we construct $G_{1},G_{2}$ for all villages across the five states to capture the attendance and adoption dynamics of people residing there. For each farmer, we compute three centrality measures for both $G_{1}$ and $G_{2}$ --- \textbf{Closeness ($CC$), Betweenness ($BC$)} and \textbf{Eigenvector ($EC$)} to account for the effect of their position in their networks on the adoption of practices. We compute all three centrality measures temporally i.e., for a farmer $f \in F$ watching the screening of a video $v$ on date $d$, we only consider the edges in temporal networks $G_{1}$ and $G_{2}$ upto date $d$. 

Next, for a video $v$ being screened to a farmer $f$ on date $d$, we consider farmers $g \in N(f,d)$ in $G_{2}$ who have adopted the video $v$ before date $d$ to measure how past adopters of the same video from the neighbourhood $N(f, d)$ of $f$ in $G_{2}$ can influence adoption of a video by farmer $f$. We measure this \textbf{Past Co-Adopter Influence (PAI)} at two levels --- the village and group of a farmer $f$ and formulate it as follows:

\begin{equation}
    PAI_{L}(f,v,d) = |\;{N(f,d) \cap A_{L}(v,d)}\;|
\end{equation}

where $L$ represents level and $A_{L}(v,d)$ denotes the adopters of video $v$ before date $d$ at level $L$. Lastly, for a farmer $f$ watching the screening of a video $v$ on date $d$, the attention given to them by the mediator during video dissemination can vary depending on the number of co-attendees $|A_{v,d}|$. We formulate this \textbf{Mediator-Farmer Ratio (MFR)} as follows: 

\begin{equation}
    MFR(f,v,d) = \frac{1}{|A_{v,d}|}
\end{equation}








\subsection{Content Features} \label{content_features}

As seen in Section~\ref{ecosystem_dataset}, Digital Green's participatory approach for video production allows them to tailor their content for local communities, making it highly specific. First, to evaluate the effectiveness of this approach, we model the \textbf{Content Specificity (CS)} of a video by leveraging information about the farmer groups it is targeted towards. If a video $v$ is targeted to a set of groups $GT(v)$, we measure the contribution of each farmer $f \in G$ where $G \in GT(v)$ as:

\begin{equation}
    CS_{f}(v) = \frac{1}{\sum_{G \in GT(v)} |G|}
\end{equation}

Second, to account for the diversity of different levels $L \in \{group,village,block,district\}$ that farmers in the targeted groups belong to, and inter-level effects in the participatory approach, we model the specificity of a video $v$ at each level $L$ as:

\begin{equation}
    CS_{L}(v) = \frac{1}{|\bigcup_{G \in GT(v)}{}{\{\bigcup_{f \in G}L(f)\}}|}
\end{equation}

where $L(f)$ denotes the identifier for the level associated with the farmer $f$. Third, to understand how the adoption of content similar to a video $v$ in the past influences its adoptions in the future, we define \textbf{Title Adoption Frequency (TA)} as: 

\begin{equation}
    TA(v, d) = \frac{\sum_{word \in Title(v)}{A(word, d)}}{S(v, d)}
\end{equation}

where $A(word, d) = $ number of adoptions of the word across all video, date pairs in a state till date $d$, and $S(v, d) = $ number of screenings of $v$ before $d$. We measure it temporally for each video, date pair ($v,d$) by computing the cumulative sum of the adoptions per word in the title of video $v$ till date $d$ and normalize it by the number of screenings of $v$ till date $d$. Fourth, to account for the farmers' attention span and understand how different duration lengths of videos help people assimilate information, we include the \textbf{duration of a video (in minutes)} as a feature. Lastly, screening of videos can be conducted during different times of the day. Thus, to consider the \textbf{time preference of farmers}, we divide these into bins of four hours each throughout the day starting 4 am -- early morning, morning, noon, evening, night, late night before encoding them as one-hot vectors.

\subsection{Demographic Features of Actors} \label{demographic_features}

Digital Green's data provides us with the information about a farmer's village and group. Therefore, to measure the extent to which the size of their community impacts their adoptions, we use their \textbf{group size} and \textbf{village size} as features. Then for each farmer $f$ viewing a video $v$ on date $d$, we measure their \textbf{active age} as the number of days between their first screening and $d$ to consider the duration of their association with DG temporally. Lastly, in Section~\ref{gender_inequalities}, we make use of the \textbf{gender} of farmers and mediators to diagnose the Digital Green ecosystem for potential inequalities. However, we refrain from using gender in our prediction model since it is a sensitive attribute.

\section{Understanding Differential Factors in Adoption} \label{differential_factors}

In this section we try to understand why some farmers adopt more videos than others based on the various factors that govern the differences between them. To study, this we define the \textbf{Adoption Rate (AR)} for each farmer as follows: 

\begin{equation}
AR(farmer) = \frac{|\;videos\;adopted\;by\;farmer\;|}{|\;videos\;viewed\;by\;farmer\;|}
\end{equation}

\begin{figure}
\centering
\begin{minipage}{.45\textwidth}
  \centering
    \includegraphics[width=0.7\linewidth]{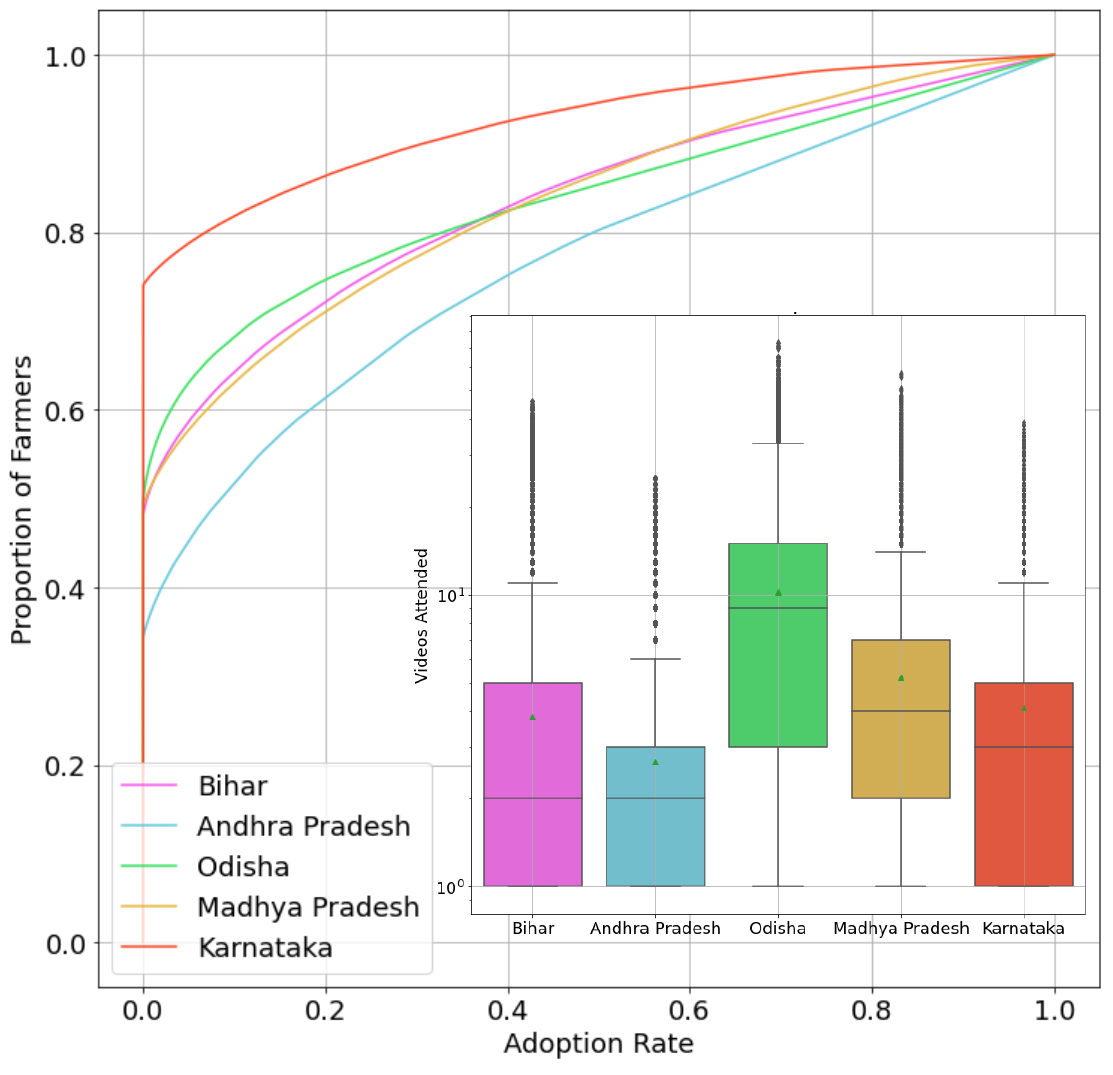}
    \caption{\textbf{CDF plot for adoption rates of farmers across the five states showing that a large percentage of Farmers have no adoptions. Inset plot represents box plot of videos attended by farmers with AR = 0.}}
    \label{fig:5}
\end{minipage}
\hfill
\begin{minipage}{.45\textwidth}
  \centering
    \includegraphics[width=1\linewidth]{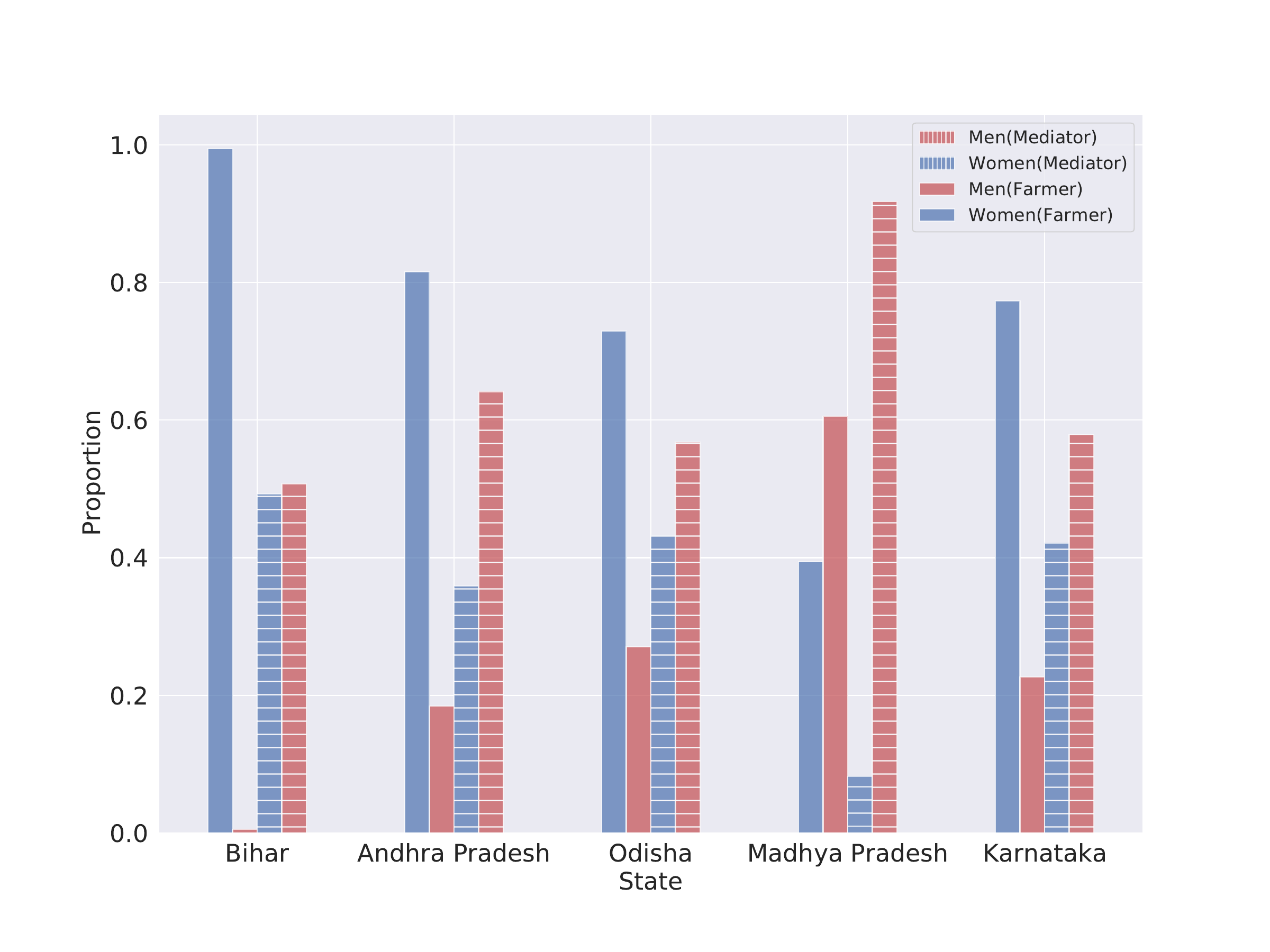}
    \caption{\textbf{Bar plot showing the proportion of farmer and mediator genders across the five states. Majority farmers in all states except Madhya Pradesh are women whereas the proportion of men is higher for mediators across all states.}}
    \label{fig:6}
\end{minipage}%
\end{figure}

We plot the CDF of adoption rate for farmers across the five states (see Figure~\ref{fig:5}). We observe that a significant percentage of the farmers have not adopted any videos - Karnataka (74\%), Odisha (49\%), Madhya Pradesh (49\%), Bihar (48\%), and Andhra Pradesh (34\%). The number of videos attended by these farmers is also very highly significantly lower than farmers who have adopted at least one video. We speculate their reasons to be attributed to their socio-economic status, lack of resources or them adopting sub-practices that do not fulfil the surveyor’s criteria for adoption. However, we plan to conduct an ethnographic investigation in the future to better understand these reasons since our data does not account for their lived experiences. For the scope of our analysis, we only consider the farmers with at least one adoption i.e., $AR > 0$.

\subsection{Why Some Farmers Adopt more than Others}\label{differnt_farmers}

We divide the farmers with $AR > 0$ across each of the five states into quartiles based on their adoption rates to understand how the lowest 25\% ($q_{1}$) and top 25\% ($q_{4}$) farmers vary in terms of factors specific to all farmer, video ($f,v$) pairs. To do so, we consider ten factors -- mediator-farmer ratio ($MFR$), content specificity ($CS$) at the farmer, village and group level, past co-adopter influence ($PAI$) at the village and group level, video duration, group size ($GS$), village size ($VS$) and active age. For the first seven factors, we consider the mean value across all videos attended by each farmer. We make use of one-tailed Welch's \textit{t}-test to evaluate our hypotheses across all the factors (Table~\ref{tab:differential_tests}). For $MFR_{\mu}$, $CS_{f\mu}$, $CS_{G\mu}$, $CS_{V\mu}$, $PAI_{G\mu}$, $PAI_{V\mu}$ and $Active\;Age$, we test the hypothesis $H_{1}: q_{1} < q_{4}$; We evaluate if higher mediator-farmer ratio, content specificity, past co-adopter influence and longer active association with DG result in higher adoption rates. For the other three, we test the hypothesis $H_{1}: q_{4} < q_{1}$ to evaluate whether longer duration of videos ($duration_{\mu}$), and larger sizes of villages ($VS$) and groups ($GS$) lead to lower adoption rates. Given that we test ten different hypotheses using the same samples, we apply the Bonferroni correction (number of measures \textit{m=10}) to the $p$ values while considering statistical significance. We only report the results for $\alpha = 0.001/m$. Mean values for $q_{1}$ and $q_{4}$ across all the ten factors are reported in the Appendix Section~\ref{mean_vals}.

\begin{table*}[!h]
  \caption{\textbf{One-Tailed Welch \textit{t}-Tests for ten factors between farmers in $q_{1}$ and $q_{4}$ of adoption rates. \textit{t}-stat is reported only in cells where $\alpha = 0.001$ after adjusting $p$ values as per the Bonferroni correction (number of measures $m=10$).}}
  \label{tab:differential_tests}
  \resizebox{\textwidth}{!}{%
  \begin{tabular}{cccccccccccc}
    \toprule
    & \multicolumn{7}{c}{\text{$q_{1} < q_{4}$}} & & \multicolumn{3}{c}{\text{$q_{4} < q_{1}$}}\\ \cline{ 2 - 8 } \cline{ 10 - 12 } 
    State & $MFR_{\mu}$& $CS_{F\mu}$& $CS_{G\mu}$ & $CS_{V\mu}$ & $PAI_{G\mu}$ & $PAI_{V\mu}$ & $Active\;Age$ && $duration_{\mu}$ & $GS$ & $VS$\\
    \midrule
    Bihar & -43.22 & -57.57& -28.31 & -30.35 & -9.91 & -51.47 & - && -12.89 & -54.89 & -23.81\\
    Andhra Pradesh & -21.96 & - & -40.30 & -40.40 & - & - & - && -6.13 & - & -21.22\\
    Odisha & - & - & -59.04 & -59.24 & -24.82 & -23.81 & - && -87.03 & - & -33.93\\
    Madhya Pradesh & -26.12 & -22.22 &-10.03 & -10.15 & -15.75 & -16.51 & - && -10.03 & -31.84  & -12.30\\
    Karnataka & - & -2.34 & -  & - & -6.64 & -3.78 & - && -20.12 & - & -35.23\\
    \bottomrule
  \end{tabular}}
\end{table*}

First, we infer that farmers with higher adoption rates ($q_{4}$) watch videos of shorter duration and belong to smaller villages across all the five states as compared to the farmers in $q_{1}$. Second, for all states except Andhra Pradesh, farmers in $q_{4}$ have significantly higher $PAI_{\mu}$ than farmers in $q_{1}$ i.e., they highly benefit from other farmers belonging to their neighbourhood in $G_{2}$ who are past adopters of a video now being screened to them. This demonstrates the role of co-adopters of a farmer in influencing adoption. Third, for all the states except Karnataka, farmers in $q_{4}$ watch videos that are more specific to their groups and villages, i.e., videos produced with the involvement of lesser groups and villages as compared to farmers in $q_{1}$. Thus, for farmers in $q_{4}$, it is likely that the participatory approach becomes more feasible when there is lesser diversity of participants in terms of groups and villages. Similarly, farmers in $q_{4}$ from Bihar, Andhra Pradesh and Karnataka also prefer watching videos that span lesser number of farmers across the targeted groups. Fourth, farmers in $q_{4}$ from Bihar, Andhra Pradesh and Madhya Pradesh watch videos with lower attendance, benefitting from a higher mediator-farmer ratio ($MFR_{\mu}$) as compared to farmers in $q_{1}$. Fifth, farmers in $q_{4}$ from Bihar and Madhya Pradesh belong to smaller groups as compared to $q_{1}$. Lastly, the mean active age of farmers in $q_{4}$ was lesser than farmers in $q_{1}$ (see Appendix Section~\ref{mean_vals}) in contrast to our hypothesis. We delve into this in more detail in Section~\ref{model_results}.

\subsection{Gender-Based Inequalities}\label{gender_inequalities}

We make use of the gender of farmers and mediators in our data to diagnose Digital Green's ecosystem for inequalities. The distribution of genders for both farmers and mediators across the five states is shown in Figure~\ref{fig:6}. We observe that majority of the farmers in all the states except Madhya Pradesh are women whereas the proportion of men is higher for mediators across all states. First, we define the adoption rate (AR) for a mediator as follows:

\begin{equation}
AR(mediator) = \frac{\sum_{v \in V(mediator)} \frac{|\;adoptions\;of\;v\;|}{|\;attendees\;of\;v\;|}}{| V (mediator) |}
\end{equation}

where $V(mediator)$ denotes the set of videos disseminated by the mediator to the farmers. We only consider attendees and adoptions for the screening conducted by the mediator. Digital Green aims to empower smallholder and marginalised farmers in villages across India, most of who are women.~\footnote{https://www.digitalgreen.org/india/} Therefore, we test the hypothesis $H_{1}: AR_{\mu}(Women) < AR_{\mu}(Men) $ for both farmers and mediators using a one-sided Welch’s \textit{t}-test across the five states to facilitate our diagnosis (Table~\ref{tab:hypo_gender}). 

\begin{table}[!htb]
    \caption{\textbf{One-Tailed Welch's \textit{t}-test for Farmer and Mediator Gender. We report \textit{t}-stat for $\alpha=$ 0.05(*), 0.001(**)}.}
    \label{tab:hypo_gender}
    \begin{subtable}{.5\linewidth}
      \centering

      \caption{One-Tailed Welch's \textit{t}-test for Farmer Gender.}
      \label{tab:hypo_persongender}
      \begin{tabular}{ccccc}
        \toprule
        \diagbox[width=8em]{\textbf{State}}{\textbf{Farmer}} & \textbf{$AR_{\mu}(M)$} & \textbf{$AR_{\mu}(W)$} & \textbf{\textit{t}-stat}\\
        \midrule
        Bihar & 0.2322 & 0.4589 &  - \\
        Andhra Pradesh & 0.6871 & 0.5737  & -60.31** \\
        Odisha & 0.7150 & 0.5207  & -57.89** \\
        Madhya Pradesh & 0.2565 & 0.2810 &  - \\
        Karnataka & 0.3712 & 0.2420 & -20.02** \\
      \bottomrule
    \end{tabular}
    \end{subtable}%
    \begin{subtable}{.5\linewidth}
      \centering

      \caption{One-Tailed Welch's \textit{t}-test for Mediator Gender.}
      \label{tab:hypo_animatorgender}
      \begin{tabular}{cccc}
        \toprule
        \diagbox[width=8em]{\textbf{State}}{\textbf{Mediator}} & \textbf{$AR_{\mu}(M)$} & \textbf{$AR_{\mu}(W)$} & \textbf{\textit{t}-stat}\\
        \midrule
        Bihar & 0.2346 & 0.2210 & -1.83* \\
        Andhra Pradesh & 0.4014 & 0.3928 & - \\
        Odisha & 0.4115 & 0.2043 & - \\
        Madhya Pradesh & 0.1566 & 0.1139 & -2.09* \\
        Karnataka & 0.0696 & 0.0862 & -\\
      \bottomrule
    \end{tabular}
    \end{subtable} 
\end{table}

For farmers, the Welch's \textit{t}-test (Table~\ref{tab:hypo_persongender}) informs that the disparities in the adoption rates of men and women are very highly significant in Andhra Pradesh, Odisha and Karnataka (greater \textit{t}-stat denotes more disparity). For mediators, the \textit{t}-test (Table~\ref{tab:hypo_animatorgender}) highlights that men are more effective mediators in Bihar and Madhya Pradesh.

\section{Predicting Adoption of Agricultural Practices}\label{predicting_adoptions}

In the previous section, we learned how farmers with lower AR ($q_{1}$) differ from those with higher AR ($q_{4}$). One of our main objectives is to help improve the adoption rates for such farmers ($q_{1}$). Therefore, to identify farmers who are less likely to adopt practices from a video screened on a particular date, we model this problem as a prediction task. The complete pipeline of adoptions, starting from preparing the content for a video, disseminating the practices via screenings and finally the farmer adopting the practice, involves several key components which we defined in Section~\ref{modelling_components}. We leverage them for our model and explain how various features impact its output differently across the five states. A prior estimate about the response of a video being screened will enable DG to support the farmers who are not likely to adopt the video and conduct on-field investigations to better understand their reasons. Therefore, we try to predict whether a farmer $f$ will adopt a video $v$ being screened to them on date $d$. The next section describes our model setup in detail.

\subsection{Model Setup:}\label{model_setup}

\begin{table}
  \caption{\textbf{Class Distribution in data across the five states.}}
  \label{tab:state_wise_training_data}
  \begin{tabular}{ccc}
    \toprule
    \diagbox[width=8em]{\textbf{State}}{\textbf{Class}} & \textbf{No Adoption} & \textbf{Adoption}\\
    \midrule
    Bihar & 2,180,336 & 1,188,162 \\
    Andhra Pradesh & 514,456 & 549,094 \\
    Odisha & 490,386  & 239,687 \\
    Madhya Pradesh & 439,639 & 149,599 \\
    Karnataka & 138,891 & 35,770 \\
  \bottomrule
\end{tabular}
\end{table}

In Section~\ref{ecosystem_dataset}, we learned that the videos of Digital Green are highly specific to the states they screen them in. In Section~\ref{differential_factors}, while understanding differential factors in adoption, we again observed most of the trends being distinct to the states. Hence, we acknowledge this diversity and divide our data across the five states based on the location where the screening is conducted. All our features have been computed temporally for each $(f,v,d)$ triplet in this timeline. Hence, we utilize a stratified split to divide the train-test data in an 80:20 ratio for our model. The distribution of both classes (No Adoption: 0, Adoption: 1) for all the $(f,v,d)$ pairs in the five states can be observed in Table~\ref{tab:state_wise_training_data}. To overcome the class imbalance, we perform down sampling for the majority class in the preparation of our train set. We train five models, one for each of the five states. We experiment three different classification techniques to predict adoptions -- (i) Logistic Regression, (ii) XGBoost \textit{(boosting stages = 25, lr = 0.1)} and (iii) Random Forest (\textit{trees = 25, depth = max}).

\subsection{Model Results and Explainability}\label{model_results}

\begin{table*}
  \caption{\textbf{Classification Results with macro-f1 and TN Rate for the three models across the five states. Random Forest produces the best results for both metrics.}}
  \label{tab:model_results}
  \begin{tabular}{ccccccccc}
    \toprule
    & \multicolumn{2}{c}{\text{Logistic Reg}} & & \multicolumn{2}{c}{\text{XGBoost}} &&
    \multicolumn{2}{c}{\text{Random Forest}}\\ \cline{ 2 - 3 } \cline{5 - 6} \cline{ 8 - 9 } 
    State & Macro-F1 & TN & & Macro-F1 & TN & & Macro-F1 & TN\\
    \midrule
    Bihar & 0.61 & 0.59 && 0.65 & 0.63 & & \textbf{0.89} & \textbf{0.90} \\
    Andhra Pradesh & 0.62 & 0.66 &&0.65& 0.63 & & \textbf{0.85} & \textbf{0.84} \\
    Odisha & 0.72 & 0.71 && 0.75 & 0.71 & & \textbf{0.87} & \textbf{0.85} \\
    Madhya Pradesh & 0.57 &  0.56 && 0.61 & 0.59 & & \textbf{0.79} & \textbf{0.79} \\
    Karnataka & 0.64 & 0.62 && 0.71 & 0.70 & & \textbf{0.82} & \textbf{0.82} \\
    \bottomrule
  \end{tabular}
\end{table*}

\begin{figure*}
    \centering
    \includegraphics[width=\textwidth]{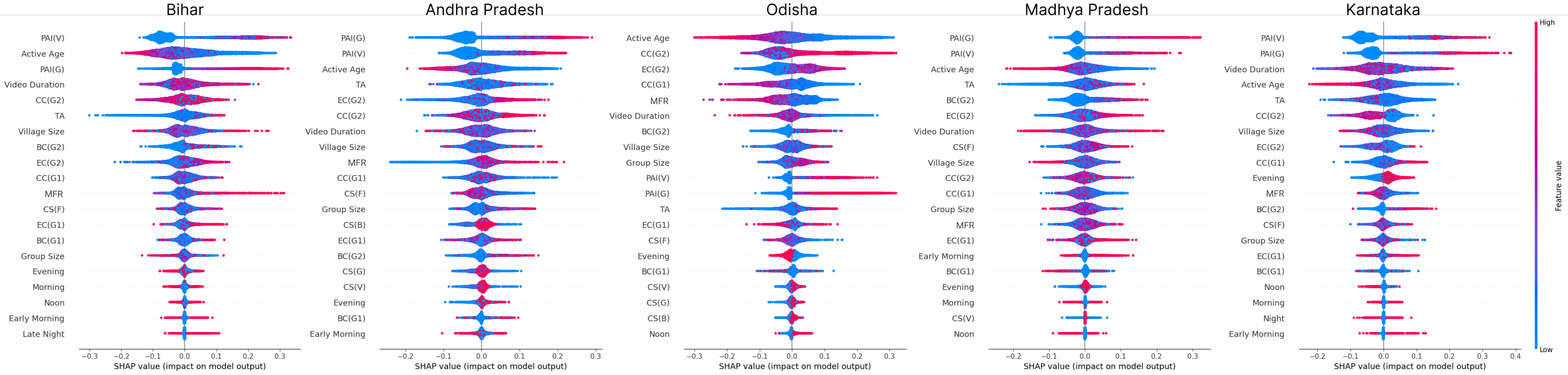}
    \caption{\textbf{SHAP summary plots for model features for all five states, with features ranked by importance from top to bottom. The x-axis represents the SHAP value for each feature. For e.g., Past Co-Adopter Influence ($PAI$) has high (red) values towards the right, indicating its positive impact on adoption (class 1) as per our model.}}
    \label{fig:shap_plots}
\end{figure*}

For evaluating our model, we use two metrics -- (i) True Negative (TN) Rate which is important in identifying farmers facing challenges in adopting practices, and (ii) the macro-f1 score to account for the class-imbalance in the dataset. Performance of all three models across the five states is summarized in Table~\ref{tab:model_results}. We find that the Random Forest Classifier outperforms the other two models on both the metrics. We use Shapley Additive Explanations, or SHAP values \cite{SHAP} to measure the feature importance. These explanations capture the contribution of each feature in the model based on local explanations \cite{SHAPlocal}. Therefore, we produce SHAP plots for  class 1 (Adoption) (see Figure~\ref{fig:shap_plots}) for every state to measure and explain the impact of different features in predicting the adoption of farming practices. 

First, we note that past co-adopter influence ($PAI$) at the group-village levels positively impacts the adoption of farming practices as per our model. This hints at the role played by past adopters of the same video who belong to the farmer’s community in transferring relevant knowledge.  Second, influential farmers (high $EC$) and farmers connecting multiple communities (high $BC$) in the co-adoption network are more likely to adopt practices from a video than others. Third, farmers from Andhra Pradesh, Odisha and Madhya Pradesh who have co-adopted with more farmers (high $CC$) have a higher likelihood of adopting practices as per our model. In conjunction with the past co-adopter influence, this indicates that a larger co-adoption neighbourhood potentially allows for more knowledge sharing. Fourth, we find that longer videos have a negative impact on adoption across all states except Bihar and Madhya Pradesh, i.e., shorter videos are preferred by the farmers. We verify this further by fitting a regression line onto the SHAP dependency plots across all states (see Fig.~\ref{fig:dep_video_duration} in Appendix Section~\ref{shap_dependency}). Fifth, for content specificity, we again supplement our inferences with the SHAP dependency plots from Appendix Section~\ref{shap_dependency}. From Fig.~\ref{fig:dep_csf}, we verify that content specificity at the farmer level ($CS_{F}$) in Bihar, Madhya Pradesh and Karnataka has a positive impact on adoptions. However, the measure of content specificity at the group, village and block levels is the same for all farmers watching a particular video, i.e., the points represented in the SHAP plots only vary for videos. $CS_{V}$ is only determined as an important feature in the SHAP plots of Andhra Pradesh, Odisha and Madhya Pradesh. Across these three states, we notice that high values of $CS_{V}$ have a very low impact on adoptions, be it positive or negative i.e., videos with high content specificity at the village level almost equally impact adoptions for all farmers. However, the same does not hold for videos with low values of $CS_{V}$; they create disproportionate outcomes for farmers in Andhra Pradesh and Madhya Pradesh. This hints that videos with low content specificity at the village level are biased towards certain villages and that it becomes difficult to employ a participatory approach when there is more diversity in villages. A similar trend can also be observed for content specificity at the group ($CS_{G}$) and block ($CS_{B}$) levels in Andhra Pradesh and Odisha. Sixth, the title adoption frequency ($TA$) positively impacts adoption in Bihar, Odisha and Madhya Pradesh i.e., videos with similar content to the previously adopted ones have a higher chance of adoption. Seventh, the active age of association of a farmer with DG negatively impacts their adoption across all the states. We plan to investigate the underlying reasons for the same in our future work. Eighth, farmers in Bihar, Andhra Pradesh and Madhya Pradesh benefit from higher mediator-farmer ratio ($MFR$) i.e., videos attended by fewer farmers in one screening are preferred. Ninth, larger group sizes have a higher likelihood of adoption in Andhra Pradesh and Odisha as opposed to Bihar and Madhya Pradesh where they negatively impact adoptions. The impact of group sizes in Karnataka is, however, mixed. Lastly, while the timing at which the video is screened plays a relatively less important role in influencing adoption, the SHAP plots give an overview of the varying timing preferences of farmers across the states, many of which are polarised.

\section{Implications} \label{implications}

Our work represents findings that can be useful for the design and evaluation of ICTs that are deployed in rural and agricultural settings. In this section, we highlight the implications of our findings from Sections~\ref{differential_factors} and~\ref{model_results}.

\begin{enumerate}
\item \textbf{Focused Assistance}: Our model accurately identifies farmers who might face challenges in adopting videos. This will enable Digital Green to assist farmers with training, equip them with resources or conduct ethnographic investigation to better understand their difficulties.

\item \textbf{Community Building}: Both statistical tests and model outcomes determine that past adopters of a video can greatly help farmers in their local community in adopting the video. Hence, we suggest building co-adoption communities at the ground level for farmers with low Past Co-Adopter Influence ($PAI$) to alleviate their adoption rates. In cases where significant farmers in a village/group have not adopted a video and their past co-adopters at the group-village levels have adopted it, a recommendation to screen that video can also be made.

\item \textbf{Recommendations for Videos}: In Bihar, Odisha and Madhya Pradesh, we suggest the use of titles for videos that have high Title Adoption Frequency ($TA$) due to its positive impact on adoptions. In cases where the videos cannot be represented by such titles due to difference in keywords, we recommend revising the video content so that it fits in a shorter duration in Odisha. We suggest the same for videos in Andhra Pradesh and Karnataka because of the negative impact of longer videos on adoptions. Further, both statistical tests and model explanations indicate that higher mediator-farmer ratio positively impacts adoptions in Bihar, Andhra Pradesh and Madhya Pradesh. Hence, we recommend improving the farmer:mediator ratio for farmers with lower adoption rates in these states.

\item \textbf{Revisiting Participatory Approach}: In the previous section, we found that low content specificity for videos at the group, village and block levels can lead to disparate outcomes for some farmers in terms of adoption. Therefore, we recommend that the participatory approach for video production spans fewer groups, villages and blocks in Andhra Pradesh, Odisha and Madhya Pradesh. However, to account for issues of agency that arise often when co-designing with marginalised groups, we also prescribe an on-ground qualitative investigation for Digital Green's participatory process across these states. An inclusive participatory approach that ensures representation across multiple axes of marginalisation including caste and gender is key in creating equal as well as equitable outcomes. 

\item \textbf{Mitigating Inequalities}: In Section~\ref{gender_inequalities}, we found significant gender-based inequalities in terms of adoption rates across three states. This will enable Digital Green in investigating and mitigating them as their service continues. 
\end{enumerate}

\section{Positionality}\label{positionality}

Our work uses computational methods for evaluating Digital Green as an ICT and developing an understanding of the disparities and various challenges faced by farmers in benefitting from it.  As researchers, while certain dimensions of our identity such as caste and gender might intersect with marginal farmers, our lived experiences most certainly do not intersect with theirs. Thus, we critically reflect on why we use certain methods, how we use them and how we address their limitations in generating findings and translating them into implications for Digital Green. 

First, participatory design is a key element in the foundation and functioning of Digital Green. However, given its epistemic differences with normative notions such as ‘generalizability’ in computational methods, we make a trade-off in terms of scale to preserve the diversity of human experiences in our dataset; we use statistical tests, build prediction models and draw inferences only at the state level instead of a country, continent or global level. Second, we draw from Abebe et. al~\cite{abebefairness} and take precautionary measures by using computational methods only as a \textit{diagnostic}, to identify and characterise challenges in the use of the ICT as a technological intervention to assist smallholder farmers. Third, we take special care in ensuring that our methods do not approximate the experiences of farmers and prescribe a qualitative inquiry to learn more about them wherever we encounter their limitations. Thus, as researchers, we also undertake the responsibility of learning more about the farmers' experiences in our future work so that any transformations in the ICT’s functioning lead to a more equal and equitable future for the farmers. Fourth, we acknowledge the unmarked experiences of people marginalized on the basis of their caste --- a marker of social stratification that actively influences how power dynamics unfold in rural areas of India. As highlighted above, we aspire to be more intersectional in our future work. Fifth, gender is collected and stored as binary in the dataset. A number of factors including but not limited to social stigamtization, lack of agency and heteronormativity continue to prevent queer and trans individuals from both urban and rural settings in India from performing their gender. Thus, it would require further research and investigation to come up with potential directions that can help change this outlook positively in both the gender and caste contexts. 
Lastly, this work is a first step in our collaboration with Digital Green. We are actively working with the DG team to qualitatively understand the experiences of the farmers, in order to better understand notions of bias (both in the social and technological sense) and how our approaches can be developed to mitigate instead of amplify these biases. 

\section{Conclusion}\label{discussion}

In this work, we looked at ten years of data from the web-based data tracker of an ICT (Digital Green) that seeks to empower rural households by enabling knowledge sharing of various types of practices. In particular, we examined the adoption of agricultural practices across five states of India -- Bihar, Andhra Pradesh, Madhya Pradesh and Karnataka. We modelled different components of the Digital Green ecosystem and used statistical methods to identify various factors that distinguish farmers with higher adoption rates from others. We diagnosed the Digital Green ecosystem to highlight gender-based inequalities among farmers in Andhra Pradesh, Odisha and Karnataka. While our analysis is currently limited to gender, we plan to include caste in our future work to investigate inequalities from an intersectional feminist lens. Next, we leveraged the modelled features and experimented with different classifiers to accurately identify farmers who might face challenges in adopting videos. We argue that this would further enable us to conduct fieldwork and ethnographic inquiry into their experiences, allowing us to account for how power dynamics unfold locally. Lastly, we explained our model results using SHAP plots and we aggregated our findings to provide implications for alleviating adoption rates of nearly a million farmers in the Digital Green ecosystem. Our research builds upon past literature on the use of ICTs across rural areas in the Global South by demonstrating the use of big data for their diagnosis and evaluation. Our work studies the role of farmer network dynamics and information specificity along with other properties of participatory video content in the adoption of diverse farming practices at a large scale. 

We acknowledge that our findings are quantitative and serve as a diagnosis for the ICT. As a result, our implications rely on qualitative fieldwork to generate experiential considerations before implementation. To this end, we plan to use a mixed-methods approach for our future work to account for on-field experiences of farmers. We will also evaluate the effectiveness of self help groups to explore whether a bottom-up approach might be more beneficial as compared to a top-down policy driven approach currently in place by the government. Finally, we aspire to expand our research to non-agricultural practices and the remaining seven states in India where Digital Green is operational.


\bibliographystyle{ACM-Reference-Format}
\bibliography{sample-base}


\begin{thebibliography}{43}


\ifx \showCODEN    \undefined \def \showCODEN     #1{\unskip}     \fi
\ifx \showDOI      \undefined \def \showDOI       #1{#1}\fi
\ifx \showISBNx    \undefined \def \showISBNx     #1{\unskip}     \fi
\ifx \showISBNxiii \undefined \def \showISBNxiii  #1{\unskip}     \fi
\ifx \showISSN     \undefined \def \showISSN      #1{\unskip}     \fi
\ifx \showLCCN     \undefined \def \showLCCN      #1{\unskip}     \fi
\ifx \shownote     \undefined \def \shownote      #1{#1}          \fi
\ifx \showarticletitle \undefined \def \showarticletitle #1{#1}   \fi
\ifx \showURL      \undefined \def \showURL       {\relax}        \fi
\providecommand\bibfield[2]{#2}
\providecommand\bibinfo[2]{#2}
\providecommand\natexlab[1]{#1}
\providecommand\showeprint[2][]{arXiv:#2}

\bibitem[Abebe et~al\mbox{.}(2020)]%
        {abebefairness}
\bibfield{author}{\bibinfo{person}{Rediet Abebe}, \bibinfo{person}{Solon
  Barocas}, \bibinfo{person}{Jon Kleinberg}, \bibinfo{person}{Karen Levy},
  \bibinfo{person}{Manish Raghavan}, {and} \bibinfo{person}{David~G.
  Robinson}.} \bibinfo{year}{2020}\natexlab{}.
\newblock \showarticletitle{Roles for Computing in Social Change}. In
  \bibinfo{booktitle}{\emph{Proceedings of the 2020 Conference on Fairness,
  Accountability, and Transparency}} (Barcelona, Spain)
  \emph{(\bibinfo{series}{FAT* '20})}. \bibinfo{publisher}{Association for
  Computing Machinery}, \bibinfo{address}{New York, NY, USA},
  \bibinfo{pages}{252–260}.
\newblock
\showISBNx{9781450369367}
\urldef\tempurl%
\url{https://doi.org/10.1145/3351095.3372871}
\showDOI{\tempurl}


\bibitem[Aker et~al\mbox{.}(2016)]%
        {promises_pitfalls_ict_agriculture}
\bibfield{author}{\bibinfo{person}{Jenny~C. Aker}, \bibinfo{person}{Ishita
  Ghosh}, {and} \bibinfo{person}{Jenna Burrell}.}
  \bibinfo{year}{2016}\natexlab{}.
\newblock \showarticletitle{The promise (and pitfalls) of ICT for agriculture
  initiatives}.
\newblock \bibinfo{journal}{\emph{Agricultural Economics}}
  \bibinfo{volume}{47}, \bibinfo{number}{S1} (\bibinfo{year}{2016}),
  \bibinfo{pages}{35--48}.
\newblock
\urldef\tempurl%
\url{https://doi.org/10.1111/agec.12301}
\showDOI{\tempurl}


\bibitem[Alcon et~al\mbox{.}(2014)]%
        {forecasting_irrigation}
\bibfield{author}{\bibinfo{person}{Francisco Alcon}, \bibinfo{person}{Sorada
  Tapsuwan}, \bibinfo{person}{José~M. Martínez-Paz}, \bibinfo{person}{Roy
  Brouwer}, {and} \bibinfo{person}{María~D. {de Miguel}}.}
  \bibinfo{year}{2014}\natexlab{}.
\newblock \showarticletitle{Forecasting deficit irrigation adoption using a
  mixed stakeholder assessment methodology}.
\newblock \bibinfo{journal}{\emph{Technological Forecasting and Social Change}}
   \bibinfo{volume}{83} (\bibinfo{year}{2014}), \bibinfo{pages}{183--193}.
\newblock
\showISSN{0040-1625}
\urldef\tempurl%
\url{https://doi.org/10.1016/j.techfore.2013.07.003}
\showDOI{\tempurl}


\bibitem[Bali et~al\mbox{.}(2013)]%
        {vkheti2}
\bibfield{author}{\bibinfo{person}{Kalika Bali}, \bibinfo{person}{Sunayana
  Sitaram}, \bibinfo{person}{Sebastien Cuendet}, {and} \bibinfo{person}{Indrani
  Medhi}.} \bibinfo{year}{2013}\natexlab{}.
\newblock \showarticletitle{A Hindi Speech Recognizer for an Agricultural Video
  Search Application}. In \bibinfo{booktitle}{\emph{Proceedings of the 3rd ACM
  Symposium on Computing for Development}} (Bangalore, India)
  \emph{(\bibinfo{series}{ACM DEV '13})}. \bibinfo{publisher}{Association for
  Computing Machinery}, \bibinfo{address}{New York, NY, USA}, Article
  \bibinfo{articleno}{5}, \bibinfo{numpages}{8}~pages.
\newblock
\showISBNx{9781450318563}
\urldef\tempurl%
\url{https://doi.org/10.1145/2442882.2442889}
\showDOI{\tempurl}


\bibitem[Barakabitze et~al\mbox{.}(2017)]%
        {barakabitze2017use}
\bibfield{author}{\bibinfo{person}{Alcardo~Alex Barakabitze},
  \bibinfo{person}{Kadeghe~G Fue}, {and} \bibinfo{person}{Camilius~Aloyce
  Sanga}.} \bibinfo{year}{2017}\natexlab{}.
\newblock \showarticletitle{The use of participatory approaches in developing
  ICT-based systems for disseminating agricultural knowledge and information
  for farmers in developing countries: The case of Tanzania}.
\newblock \bibinfo{journal}{\emph{The Electronic Journal of Information Systems
  in Developing Countries}} \bibinfo{volume}{78}, \bibinfo{number}{1}
  (\bibinfo{year}{2017}), \bibinfo{pages}{1--23}.
\newblock


\bibitem[Bede et~al\mbox{.}(2021)]%
        {videomediatedrural}
\bibfield{author}{\bibinfo{person}{Lauréano Bede}, \bibinfo{person}{Florent
  Okry}, {and} \bibinfo{person}{Simplice~D. Vodouhe}.}
  \bibinfo{year}{2021}\natexlab{}.
\newblock \showarticletitle{Video-mediated rural learning: effects of images
  and languages on farmers’ learning in Benin Republic}.
\newblock \bibinfo{journal}{\emph{Development in Practice}}
  \bibinfo{volume}{31}, \bibinfo{number}{1} (\bibinfo{year}{2021}),
  \bibinfo{pages}{59--68}.
\newblock
\urldef\tempurl%
\url{https://doi.org/10.1080/09614524.2020.1788508}
\showDOI{\tempurl}
\showeprint{https://doi.org/10.1080/09614524.2020.1788508}


\bibitem[Bello-Bravo et~al\mbox{.}(2018)]%
        {educationalanimated}
\bibfield{author}{\bibinfo{person}{Julia Bello-Bravo}, \bibinfo{person}{Manuele
  Tamò}, \bibinfo{person}{Elie~Ayitondji Dannon}, {and}
  \bibinfo{person}{Barry~Robert Pittendrigh}.} \bibinfo{year}{2018}\natexlab{}.
\newblock \showarticletitle{An assessment of learning gains from educational
  animated videos versus traditional extension presentations among farmers in
  Benin}.
\newblock \bibinfo{journal}{\emph{Information Technology for Development}}
  \bibinfo{volume}{24}, \bibinfo{number}{2} (\bibinfo{year}{2018}),
  \bibinfo{pages}{224--244}.
\newblock
\urldef\tempurl%
\url{https://doi.org/10.1080/02681102.2017.1298077}
\showDOI{\tempurl}
\showeprint{https://doi.org/10.1080/02681102.2017.1298077}


\bibitem[Bentley et~al\mbox{.}(2019)]%
        {bentlay}
\bibfield{author}{\bibinfo{person}{Jeffery~W. Bentley},
  \bibinfo{person}{Paul~Van Mele}, \bibinfo{person}{Nafissath~Fousseni Barres},
  \bibinfo{person}{Florent Okry}, {and} \bibinfo{person}{Jonas Wanvoeke}.}
  \bibinfo{year}{2019}\natexlab{}.
\newblock \showarticletitle{Smallholders download and share videos from the
  Internet to learn about sustainable agriculture}.
\newblock \bibinfo{journal}{\emph{International Journal of Agricultural
  Sustainability}} \bibinfo{volume}{17}, \bibinfo{number}{1}
  (\bibinfo{year}{2019}), \bibinfo{pages}{92--107}.
\newblock
\urldef\tempurl%
\url{https://doi.org/10.1080/14735903.2019.1567246}
\showDOI{\tempurl}
\showeprint{https://doi.org/10.1080/14735903.2019.1567246}


\bibitem[Bruges and Smith(2008)]%
        {bruges2008participatory}
\bibfield{author}{\bibinfo{person}{Murray Bruges} {and} \bibinfo{person}{Willie
  Smith}.} \bibinfo{year}{2008}\natexlab{}.
\newblock \showarticletitle{Participatory approaches for sustainable
  agriculture: A contradiction in terms?}
\newblock \bibinfo{journal}{\emph{Agriculture and Human Values}}
  \bibinfo{volume}{25}, \bibinfo{number}{1} (\bibinfo{year}{2008}),
  \bibinfo{pages}{13--23}.
\newblock


\bibitem[Chambers and Thrupp(1994)]%
        {chambers1994farmer}
\bibfield{author}{\bibinfo{person}{Robert Chambers} {and}
  \bibinfo{person}{Lori~Ann Thrupp}.} \bibinfo{year}{1994}\natexlab{}.
\newblock \bibinfo{booktitle}{\emph{Farmer first: farmer innovation and
  agricultural research}}.
\newblock \bibinfo{publisher}{Karthala Editions}.
\newblock


\bibitem[Conley and Udry(2001)]%
        {social_networks_ghana}
\bibfield{author}{\bibinfo{person}{Timothy Conley} {and}
  \bibinfo{person}{Christopher Udry}.} \bibinfo{year}{2001}\natexlab{}.
\newblock \showarticletitle{Social Learning through Networks: The Adoption of
  New Agricultural Technologies in Ghana}.
\newblock \bibinfo{journal}{\emph{American Journal of Agricultural Economics}}
  \bibinfo{volume}{83}, \bibinfo{number}{3} (\bibinfo{year}{2001}),
  \bibinfo{pages}{668--673}.
\newblock
\showISSN{00029092, 14678276}
\urldef\tempurl%
\url{http://www.jstor.org/stable/1245097}
\showURL{%
\tempurl}


\bibitem[Cu{\'e}llar-Padilla and Calle-Collado(2011)]%
        {cuellar2011can}
\bibfield{author}{\bibinfo{person}{Mamen Cu{\'e}llar-Padilla} {and}
  \bibinfo{person}{{\'A}ngel Calle-Collado}.} \bibinfo{year}{2011}\natexlab{}.
\newblock \showarticletitle{Can we find solutions with people? Participatory
  action research with small organic producers in Andalusia}.
\newblock \bibinfo{journal}{\emph{Journal of Rural Studies}}
  \bibinfo{volume}{27}, \bibinfo{number}{4} (\bibinfo{year}{2011}),
  \bibinfo{pages}{372--383}.
\newblock


\bibitem[Cuendet et~al\mbox{.}(2013)]%
        {vkheti1}
\bibfield{author}{\bibinfo{person}{Sebastien Cuendet}, \bibinfo{person}{Indrani
  Medhi}, \bibinfo{person}{Kalika Bali}, {and} \bibinfo{person}{Edward
  Cutrell}.} \bibinfo{year}{2013}\natexlab{}.
\newblock \bibinfo{booktitle}{\emph{VideoKheti: Making Video Content Accessible
  to Low-Literate and Novice Users}}.
\newblock \bibinfo{publisher}{Association for Computing Machinery},
  \bibinfo{address}{New York, NY, USA}, \bibinfo{pages}{2833–2842}.
\newblock
\showISBNx{9781450318990}
\urldef\tempurl%
\url{https://doi.org/10.1145/2470654.2481392}
\showURL{%
\tempurl}


\bibitem[Feder and Umali(1993)]%
        {adoption_agriculture_innovations}
\bibfield{author}{\bibinfo{person}{Gershon Feder} {and}
  \bibinfo{person}{Dina~L. Umali}.} \bibinfo{year}{1993}\natexlab{}.
\newblock \showarticletitle{The adoption of agricultural innovations: A
  review}.
\newblock \bibinfo{journal}{\emph{Technological Forecasting and Social Change}}
  \bibinfo{volume}{43}, \bibinfo{number}{3} (\bibinfo{year}{1993}),
  \bibinfo{pages}{215--239}.
\newblock
\showISSN{0040-1625}
\urldef\tempurl%
\url{https://doi.org/10.1016/0040-1625(93)90053-A}
\showDOI{\tempurl}
\newblock
\shownote{Special Issue Technology and Innovation In Agriculture and Natural
  Resources}.


\bibitem[Ford and Moore(2021)]%
        {farmingvlogs}
\bibfield{author}{\bibinfo{person}{Hannah Ford} {and} \bibinfo{person}{Austin
  Moore}.} \bibinfo{year}{2021}\natexlab{}.
\newblock \showarticletitle{Framing Farmers: A Content Analysis of YouTube
  Farming Vlogs}. In \bibinfo{booktitle}{\emph{Conference: 2021 National AAAE
  Conference - Virtual (The American Association for Agricultural Education)}}.
\newblock
\urldef\tempurl%
\url{https://doi.org/10.13140/RG.2.2.16916.83840}
\showDOI{\tempurl}


\bibitem[Foundation(2017)]%
        {ibef}
\bibfield{author}{\bibinfo{person}{India Brand~Equity Foundation}.}
  \bibinfo{year}{2017}\natexlab{}.
\newblock \bibinfo{title}{Analysis about Indian agriculture industry, market
  size, Export \&amp; Investment Opportunity}.
\newblock
\newblock
\urldef\tempurl%
\url{https://www.ibef.org/industry/agriculture-india.aspx}
\showURL{%
\tempurl}


\bibitem[Gandhi et~al\mbox{.}(2007)]%
        {Gandhi}
\bibfield{author}{\bibinfo{person}{Rikin Gandhi}, \bibinfo{person}{Rajesh
  Veeraraghavan}, \bibinfo{person}{Kentaro Toyama}, {and}
  \bibinfo{person}{Vanaja Ramprasad}.} \bibinfo{year}{2007}\natexlab{}.
\newblock \showarticletitle{Digital Green: Participatory video for agricultural
  extension}. In \bibinfo{booktitle}{\emph{2007 International Conference on
  Information and Communication Technologies and Development}}.
  \bibinfo{pages}{1--10}.
\newblock
\urldef\tempurl%
\url{https://doi.org/10.1109/ICTD.2007.4937388}
\showDOI{\tempurl}


\bibitem[Heeks(2009)]%
        {ict4d2}
\bibfield{author}{\bibinfo{person}{Richard Heeks}.}
  \bibinfo{year}{2009}\natexlab{}.
\newblock \showarticletitle{The ICT4D 2.0 manifesto: Where next for ICTs and
  international development?}
\newblock \bibinfo{journal}{\emph{Development Informatics Working Paper}}
  \bibinfo{number}{42} (\bibinfo{year}{2009}).
\newblock


\bibitem[Hinz et~al\mbox{.}(2020)]%
        {land_use_socioeconomic}
\bibfield{author}{\bibinfo{person}{R. Hinz}, \bibinfo{person}{T.~B. Sulser},
  \bibinfo{person}{R. Huefner}, \bibinfo{person}{D. Mason-D’Croz},
  \bibinfo{person}{S. Dunston}, \bibinfo{person}{S. Nautiyal},
  \bibinfo{person}{C. Ringler}, \bibinfo{person}{J. Schuengel},
  \bibinfo{person}{P. Tikhile}, \bibinfo{person}{F. Wimmer}, {and}
  \bibinfo{person}{R. Schaldach}.} \bibinfo{year}{2020}\natexlab{}.
\newblock \showarticletitle{Agricultural Development and Land Use Change in
  India: A Scenario Analysis of Trade-Offs Between UN Sustainable Development
  Goals (SDGs)}.
\newblock \bibinfo{journal}{\emph{Earth's Future}} \bibinfo{volume}{8},
  \bibinfo{number}{2} (\bibinfo{year}{2020}).
\newblock
\urldef\tempurl%
\url{https://doi.org/10.1029/2019EF001287}
\showDOI{\tempurl}


\bibitem[Kamilaris et~al\mbox{.}(2017)]%
        {big_data_analytics_review}
\bibfield{author}{\bibinfo{person}{Andreas Kamilaris}, \bibinfo{person}{Andreas
  Kartakoullis}, {and} \bibinfo{person}{Francesc~X. Prenafeta-Boldú}.}
  \bibinfo{year}{2017}\natexlab{}.
\newblock \showarticletitle{A review on the practice of big data analysis in
  agriculture}.
\newblock \bibinfo{journal}{\emph{Computers and Electronics in Agriculture}}
  \bibinfo{volume}{143} (\bibinfo{year}{2017}), \bibinfo{pages}{23--37}.
\newblock
\showISSN{0168-1699}
\urldef\tempurl%
\url{https://doi.org/10.1016/j.compag.2017.09.037}
\showDOI{\tempurl}


\bibitem[Karubanga et~al\mbox{.}(2017)]%
        {karubanga2017farmer}
\bibfield{author}{\bibinfo{person}{Gabriel Karubanga}, \bibinfo{person}{P
  Kibwika}, \bibinfo{person}{Florent Okry}, {and} \bibinfo{person}{H Sseguya}.}
  \bibinfo{year}{2017}\natexlab{}.
\newblock \showarticletitle{How farmer videos trigger social learning to
  enhance innovation among smallholder rice farmers in Uganda}.
\newblock \bibinfo{journal}{\emph{Cogent Food \& Agriculture}}
  \bibinfo{volume}{3}, \bibinfo{number}{1} (\bibinfo{year}{2017}),
  \bibinfo{pages}{1368105}.
\newblock


\bibitem[Knoche et~al\mbox{.}(2010)]%
        {knoche2010missing}
\bibfield{author}{\bibinfo{person}{Hendrik Knoche}, \bibinfo{person}{PRS Rao},
  {and} \bibinfo{person}{J Huang}.} \bibinfo{year}{2010}\natexlab{}.
\newblock \showarticletitle{The missing H in ICTD: Lessons learned from the
  development of an agricultural decision support tool: Mobile HCI and
  Technical ICTD: A Methodological Perspective Workshop}.
\newblock  (\bibinfo{year}{2010}).
\newblock


\bibitem[Knowler and Bradshaw(2007)]%
        {farmer_adoption_review_synthesis}
\bibfield{author}{\bibinfo{person}{Duncan Knowler} {and} \bibinfo{person}{Ben
  Bradshaw}.} \bibinfo{year}{2007}\natexlab{}.
\newblock \showarticletitle{Farmers’ adoption of conservation agriculture: A
  review and synthesis of recent research}.
\newblock \bibinfo{journal}{\emph{Food Policy}} \bibinfo{volume}{32},
  \bibinfo{number}{1} (\bibinfo{year}{2007}), \bibinfo{pages}{25--48}.
\newblock
\showISSN{0306-9192}
\urldef\tempurl%
\url{https://doi.org/10.1016/j.foodpol.2006.01.003}
\showDOI{\tempurl}


\bibitem[Lin et~al\mbox{.}(2021)]%
        {farmers_social_networks_myanmar}
\bibfield{author}{\bibinfo{person}{Tian Lin}, \bibinfo{person}{Aung~Phyo Ko},
  \bibinfo{person}{Maung~Maung Than}, \bibinfo{person}{Delia~C. Catacutan},
  \bibinfo{person}{Robert~F. Finlayson}, {and} \bibinfo{person}{Marney~E.
  Isaac}.} \bibinfo{year}{2021}\natexlab{}.
\newblock \showarticletitle{Farmer social networks: The role of advice ties and
  organizational leadership in agroforestry adoption}.
\newblock \bibinfo{journal}{\emph{PLOS ONE}} \bibinfo{volume}{16},
  \bibinfo{number}{8} (\bibinfo{date}{08} \bibinfo{year}{2021}),
  \bibinfo{pages}{1--18}.
\newblock
\urldef\tempurl%
\url{https://doi.org/10.1371/journal.pone.0255987}
\showDOI{\tempurl}


\bibitem[Lundberg and Lee(2017)]%
        {SHAP}
\bibfield{author}{\bibinfo{person}{Scott~M. Lundberg} {and}
  \bibinfo{person}{Su-In Lee}.} \bibinfo{year}{2017}\natexlab{}.
\newblock \showarticletitle{A Unified Approach to Interpreting Model
  Predictions}. In \bibinfo{booktitle}{\emph{Proceedings of the 31st
  International Conference on Neural Information Processing Systems}} (Long
  Beach, California, USA) \emph{(\bibinfo{series}{NIPS'17})}.
  \bibinfo{publisher}{Curran Associates Inc.}, \bibinfo{address}{Red Hook, NY,
  USA}, \bibinfo{pages}{4768–4777}.
\newblock
\showISBNx{9781510860964}


\bibitem[Lynch and Gregor(2004)]%
        {lynch2004user}
\bibfield{author}{\bibinfo{person}{Teresa Lynch} {and} \bibinfo{person}{Shirley
  Gregor}.} \bibinfo{year}{2004}\natexlab{}.
\newblock \showarticletitle{User participation in decision support systems
  development: influencing system outcomes}.
\newblock \bibinfo{journal}{\emph{European Journal of Information Systems}}
  \bibinfo{volume}{13}, \bibinfo{number}{4} (\bibinfo{year}{2004}),
  \bibinfo{pages}{286--301}.
\newblock


\bibitem[Maertens and Barrett(2013)]%
        {measuring_social_networks}
\bibfield{author}{\bibinfo{person}{Annemie Maertens} {and}
  \bibinfo{person}{Christopher~B. Barrett}.} \bibinfo{year}{2013}\natexlab{}.
\newblock \showarticletitle{Measuring Social Networks' Effects on Agricultural
  Technology Adoption}.
\newblock \bibinfo{journal}{\emph{American Journal of Agricultural Economics}}
  \bibinfo{volume}{95}, \bibinfo{number}{2} (\bibinfo{year}{2013}),
  \bibinfo{pages}{353--359}.
\newblock
\showISSN{00029092, 14678276}
\urldef\tempurl%
\url{http://www.jstor.org/stable/23358403}
\showURL{%
\tempurl}


\bibitem[Maina et~al\mbox{.}(2020)]%
        {socioeconomic_determinants_grass_kenya}
\bibfield{author}{\bibinfo{person}{K.~W. Maina}, \bibinfo{person}{C.~N. Ritho},
  \bibinfo{person}{B.~A. Lukuyu}, {and} \bibinfo{person}{E.~J.~O. Rao}.}
  \bibinfo{year}{2020}\natexlab{}.
\newblock \showarticletitle{Socio-economic determinants and impact of adopting
  climate-smart Brachiaria grass among dairy farmers in Eastern and Western
  regions of Kenya}.
\newblock \bibinfo{journal}{\emph{Heliyon}} \bibinfo{volume}{6},
  \bibinfo{number}{6} (\bibinfo{date}{29 Jun} \bibinfo{year}{2020}),
  \bibinfo{pages}{e04335--e04335}.
\newblock
\showISSN{2405-8440}
\urldef\tempurl%
\url{https://doi.org/10.1016/j.heliyon.2020.e04335}
\showDOI{\tempurl}


\bibitem[Mittal and Mehar(2012)]%
        {mittalagricultural_informationnetworks}
\bibfield{author}{\bibinfo{person}{S. Mittal} {and} \bibinfo{person}{M.
  Mehar}.} \bibinfo{year}{2012}\natexlab{}.
\newblock \bibinfo{booktitle}{\emph{Agricultural information networks,
  information needs and risk management strategies: A survey of farmers in
  Indo-Gangetic Plains of India}}.
\newblock \bibinfo{publisher}{CIMMYT}.
\newblock
\showISBNx{9786078263318}
\urldef\tempurl%
\url{https://books.google.com.kw/books?id=RIugAgAAQBAJ}
\showURL{%
\tempurl}


\bibitem[Mittal and Mehar(2016)]%
        {socioeconomic_factors_india}
\bibfield{author}{\bibinfo{person}{Surabhi Mittal} {and} \bibinfo{person}{Mamta
  Mehar}.} \bibinfo{year}{2016}\natexlab{}.
\newblock \showarticletitle{Socio-economic Factors Affecting Adoption of Modern
  Information and Communication Technology by Farmers in India: Analysis Using
  Multivariate Probit Model}.
\newblock \bibinfo{journal}{\emph{The Journal of Agricultural Education and
  Extension}} \bibinfo{volume}{22}, \bibinfo{number}{2} (\bibinfo{year}{2016}),
  \bibinfo{pages}{199--212}.
\newblock
\urldef\tempurl%
\url{https://doi.org/10.1080/1389224X.2014.997255}
\showDOI{\tempurl}
\showeprint{https://doi.org/10.1080/1389224X.2014.997255}


\bibitem[Mwangi and Kariuki(2015)]%
        {categorisation_factors_adoptions}
\bibfield{author}{\bibinfo{person}{Margaret~R Mwangi} {and}
  \bibinfo{person}{Sam Kariuki}.} \bibinfo{year}{2015}\natexlab{}.
\newblock \showarticletitle{Factors Determining Adoption of New Agricultural
  Technology by Smallholder Farmers in Developing Countries}.
\newblock \bibinfo{journal}{\emph{Journal of economics and sustainable
  development}}  \bibinfo{volume}{6} (\bibinfo{year}{2015}),
  \bibinfo{pages}{208--216}.
\newblock


\bibitem[Pagliarino et~al\mbox{.}(2020)]%
        {pagliarino2020participatory}
\bibfield{author}{\bibinfo{person}{Elena Pagliarino},
  \bibinfo{person}{Francesca Orlando}, \bibinfo{person}{Valentina Vaglia},
  \bibinfo{person}{Secondo Rolfo}, {and} \bibinfo{person}{Stefano Bocchi}.}
  \bibinfo{year}{2020}\natexlab{}.
\newblock \showarticletitle{Participatory research for sustainable agriculture:
  the case of the Italian agroecological rice network}.
\newblock \bibinfo{journal}{\emph{European Journal of Futures Research}}
  \bibinfo{volume}{8}, \bibinfo{number}{1} (\bibinfo{year}{2020}),
  \bibinfo{pages}{1--16}.
\newblock


\bibitem[Pannell et~al\mbox{.}(2006)]%
        {promoting_adoption}
\bibfield{author}{\bibinfo{person}{D.~J. Pannell}, \bibinfo{person}{G.~R.
  Marshall}, \bibinfo{person}{N. Barr}, \bibinfo{person}{A. Curtis},
  \bibinfo{person}{F. Vanclay}, {and} \bibinfo{person}{R. Wilkinson}.}
  \bibinfo{year}{2006}\natexlab{}.
\newblock \showarticletitle{Understanding and promoting adoption of
  conservation practices by rural landholders}.
\newblock \bibinfo{journal}{\emph{Australian Journal of Experimental
  Agriculture}} \bibinfo{volume}{46}, \bibinfo{number}{11}
  (\bibinfo{year}{2006}), \bibinfo{pages}{1407--1424}.
\newblock
\urldef\tempurl%
\url{https://doi.org/10.1071/EA05037}
\showDOI{\tempurl}


\bibitem[Parikh et~al\mbox{.}(2007)]%
        {4937421}
\bibfield{author}{\bibinfo{person}{Tapan~S. Parikh}, \bibinfo{person}{Neil
  Patel}, {and} \bibinfo{person}{Yael Schwartzman}.}
  \bibinfo{year}{2007}\natexlab{}.
\newblock \showarticletitle{A survey of information systems reaching small
  producers in global agricultural value chains}. In
  \bibinfo{booktitle}{\emph{2007 International Conference on Information and
  Communication Technologies and Development}}. \bibinfo{pages}{1--11}.
\newblock
\urldef\tempurl%
\url{https://doi.org/10.1109/ICTD.2007.4937421}
\showDOI{\tempurl}


\bibitem[Patel et~al\mbox{.}(2006)]%
        {ishakti}
\bibfield{author}{\bibinfo{person}{S. Patel}, \bibinfo{person}{O. Bataveljic},
  \bibinfo{person}{P.J. Lisboa}, \bibinfo{person}{C. Hawkins}, {and}
  \bibinfo{person}{R. Rajan}.} \bibinfo{year}{2006}\natexlab{}.
\newblock \showarticletitle{iShakti--Crossing the Digital Divide in Rural
  India}. In \bibinfo{booktitle}{\emph{2006 IEEE/WIC/ACM International
  Conference on Web Intelligence (WI 2006 Main Conference
  Proceedings)(WI'06)}}. \bibinfo{pages}{1061--1065}.
\newblock
\urldef\tempurl%
\url{https://doi.org/10.1109/WI.2006.107}
\showDOI{\tempurl}


\bibitem[Ramachandran et~al\mbox{.}(2010)]%
        {ASHA}
\bibfield{author}{\bibinfo{person}{Divya Ramachandran}, \bibinfo{person}{John
  Canny}, \bibinfo{person}{Prabhu~Dutta Das}, {and} \bibinfo{person}{Edward
  Cutrell}.} \bibinfo{year}{2010}\natexlab{}.
\newblock \bibinfo{booktitle}{\emph{Mobile-Izing Health Workers in Rural
  India}}.
\newblock \bibinfo{publisher}{Association for Computing Machinery},
  \bibinfo{address}{New York, NY, USA}, \bibinfo{pages}{1889–1898}.
\newblock
\showISBNx{9781605589299}
\urldef\tempurl%
\url{https://doi.org/10.1145/1753326.1753610}
\showURL{%
\tempurl}


\bibitem[Ribeiro et~al\mbox{.}(2016)]%
        {SHAPlocal}
\bibfield{author}{\bibinfo{person}{Marco~Tulio Ribeiro},
  \bibinfo{person}{Sameer Singh}, {and} \bibinfo{person}{Carlos Guestrin}.}
  \bibinfo{year}{2016}\natexlab{}.
\newblock \showarticletitle{"Why Should I Trust You?": Explaining the
  Predictions of Any Classifier}. In \bibinfo{booktitle}{\emph{Proceedings of
  the 22nd ACM SIGKDD International Conference on Knowledge Discovery and Data
  Mining}} (San Francisco, California, USA) \emph{(\bibinfo{series}{KDD '16})}.
  \bibinfo{publisher}{Association for Computing Machinery},
  \bibinfo{address}{New York, NY, USA}, \bibinfo{pages}{1135–1144}.
\newblock
\showISBNx{9781450342322}
\urldef\tempurl%
\url{https://doi.org/10.1145/2939672.2939778}
\showDOI{\tempurl}


\bibitem[Rose et~al\mbox{.}(2021)]%
        {rose2021videos}
\bibfield{author}{\bibinfo{person}{DC Rose}, \bibinfo{person}{CA Chivers},
  \bibinfo{person}{A de Boon}, \bibinfo{person}{J Schillings},
  \bibinfo{person}{R Smith}, \bibinfo{person}{L Lishman}, {and}
  \bibinfo{person}{K Bliss}.} \bibinfo{year}{2021}\natexlab{}.
\newblock \showarticletitle{Videos and podcasts as potential approaches for
  knowledge exchange with farmers: testing their potential role in ELM}.
\newblock  (\bibinfo{year}{2021}).
\newblock


\bibitem[Sawant et~al\mbox{.}(2016)]%
        {OrganizedDA_pride_model}
\bibfield{author}{\bibinfo{person}{Minal Sawant}, \bibinfo{person}{Rajesh
  Urkude}, {and} \bibinfo{person}{Sandip Jawale}.}
  \bibinfo{year}{2016}\natexlab{}.
\newblock \showarticletitle{Organized Data and Information for Efficacious
  Agriculture Using PRIDE{\texttrademark} Model}.
\newblock \bibinfo{journal}{\emph{The International Food and Agribusiness
  Management Review}}  \bibinfo{volume}{19} (\bibinfo{year}{2016}),
  \bibinfo{pages}{115--130}.
\newblock


\bibitem[Stone(2007)]%
        {stone2007agricultural}
\bibfield{author}{\bibinfo{person}{Glenn~Davis Stone}.}
  \bibinfo{year}{2007}\natexlab{}.
\newblock \showarticletitle{Agricultural deskilling and the spread of
  genetically modified cotton in Warangal}.
\newblock \bibinfo{journal}{\emph{Current anthropology}} \bibinfo{volume}{48},
  \bibinfo{number}{1} (\bibinfo{year}{2007}), \bibinfo{pages}{67--103}.
\newblock


\bibitem[Tewathia et~al\mbox{.}(2020)]%
        {social_inequalities_india}
\bibfield{author}{\bibinfo{person}{Nidhi Tewathia}, \bibinfo{person}{Anant
  Kamath}, {and} \bibinfo{person}{P.~Vigneswara Ilavarasan}.}
  \bibinfo{year}{2020}\natexlab{}.
\newblock \showarticletitle{Social inequalities, fundamental inequities, and
  recurring of the digital divide: Insights from India}.
\newblock \bibinfo{journal}{\emph{Technology in Society}}  \bibinfo{volume}{61}
  (\bibinfo{year}{2020}), \bibinfo{pages}{101251}.
\newblock
\showISSN{0160-791X}
\urldef\tempurl%
\url{https://doi.org/10.1016/j.techsoc.2020.101251}
\showDOI{\tempurl}


\bibitem[Vangala et~al\mbox{.}(2015)]%
        {vangala2015icts}
\bibfield{author}{\bibinfo{person}{Ram Naresh~Kumar Vangala},
  \bibinfo{person}{Maitrayee Mukerji}, {and} \bibinfo{person}{BN Hiremath}.}
  \bibinfo{year}{2015}\natexlab{}.
\newblock \showarticletitle{ICTs for agriculture knowledge management: insights
  from DHRUVA, India}. In \bibinfo{booktitle}{\emph{Proceedings of the Seventh
  International Conference on Information and Communication Technologies and
  Development}}. \bibinfo{pages}{1--4}.
\newblock


\bibitem[Zossou et~al\mbox{.}(2012)]%
        {zossou2012participatory}
\bibfield{author}{\bibinfo{person}{Esperance Zossou}, \bibinfo{person}{Paul
  Van~Mele}, \bibinfo{person}{Jonas Wanvoeke}, {and} \bibinfo{person}{Philippe
  Lebailly}.} \bibinfo{year}{2012}\natexlab{}.
\newblock \showarticletitle{Participatory impact assessment of rice parboiling
  videos with women in Benin}.
\newblock \bibinfo{journal}{\emph{Experimental Agriculture}}
  \bibinfo{volume}{48}, \bibinfo{number}{3} (\bibinfo{year}{2012}),
  \bibinfo{pages}{438--447}.
\newblock


\end{thebibliography}



\begin{appendices}

\section{Mean Values for Table~\ref{tab:differential_tests}} \label{mean_vals}
Mean values for both $q_{1}$ and $q_{4}$ across all the ten factors corresponding to the One-Tailed Welch's \textit{t}-test are given in Table~\ref{tab:meanvals}, grouped by hypothesis.

\begin{table*}[hp]
    \caption{\textbf{Mean Values corresponding to One-Tailed Welch \textit{t}-Tests for ten factors between farmers in $q_{1}$ and $q_{4}$ from Table~\ref{tab:differential_tests}: $MFR_{\mu}$, $CS_{f\mu}$, $CS_{G\mu}$, $CS_{V\mu}$, $PAI_{G\mu}$, $PAI_{V\mu}$, $Active\;Age$, $duration_{\mu}$, $GS$ and $VS$}}
    \label{tab:meanvals}
    \begin{subtable}{0.9\textwidth}
    \centering

      \caption{$H_{1}: q_{1} < q_{4}$}
      \label{tab:meanvalsq1q4}
      \resizebox{\textwidth}{!}{%
      \begin{tabular}{ccccccccccccccccccccccc}
    \toprule
    & \multicolumn{2}{c}{\text{$MFR_{\mu}$}} & &
    \multicolumn{2}{c}{\text{$CS_{F\mu}$}} & &
    \multicolumn{2}{c}{\text{$CS_{G\mu}$}} & &
    \multicolumn{2}{c}{\text{$CS_{V\mu}$}} & & \multicolumn{2}{c}{\text{$PAI_{G\mu}$}} & & \multicolumn{2}{c}{\text{$PAI_{V\mu}$}} && 
    \multicolumn{2}{c}{\text{$Active\;Age$}}\\ \cline{ 2 - 3 } \cline{ 5 - 6 } \cline{ 8 - 9 } \cline{ 11 - 12 } \cline{14-15} \cline{ 17-18 } \cline{20-21}
    State & $q_{1}$ & $q_{4}$ && $q_{1}$ & $q_{4}$ && $q_{1}$ & $q_{4}$ && $q_{1}$ & $q_{4}$ && $q_{1}$ & $q_{4}$ && $q_{1}$ & $q_{4}$ && $q_{1}$ & $q_{4}$\\
    \midrule
    Bihar & 0.048 & 0.052 &&0.049 & 0.045 && 0.560& 0.540&& 0.560 & 0.539 && 0.448 & 0.519 && 8.21 & 13.41 && 637.53 & 191.52 \\
    Andhra Pradesh & 0.060 & 0.066 && 0.052 & 0.051 && 0.817 & 0.739 && 0.817 & 0.739 && 4.33 & 2.56 && 17.58 & 13.45 && 266.26 & 57.66\\
    Odisha & 0.061 & 0.046 && 0.048 & 0.044 && 0.901&0.759 && 0.901 & 0.757 && 0.279 & 0.690 && 2.87 & 5.32 && 915.81 & 276.22\\
    Madhya Pradesh & 0.069 & 0.077 && 0.066 & 0.058 && 0.974&0.956 && 0.964 & 0.943 && 0.396 & 0.625 && 2.29 & 3.31 && 506.48 & 323.43\\
    Karnataka & 0.099 & 0.090 && 0.084 & 0.080 && 0.980&0.965 && 0.971 & 0.959 && 0.387 & 0.587 && 4.81 & 5.83 && 664.18 & 178.07\\
    \bottomrule
  \end{tabular}}
    \end{subtable}
    \newline
    \newline
    \begin{subtable}{0.9\textwidth}
    \centering

      \caption{$H_{1}: q_{4} < q_{1}$}
      \label{tab:meanvalsq4q1}
      \begin{tabular}{ccccccccc}
    \toprule
    & \multicolumn{2}{c}{\text{$duration_{\mu}$}} & & \multicolumn{2}{c}{\text{$GS$}} & & \multicolumn{2}{c}{\text{$VS$}} \\ \cline{ 2 - 3 } \cline{ 5 - 6 } \cline{ 8 - 9 }
    State & $q_{1}$ & $q_{4}$ && $q_{1}$ & $q_{4}$ && $q_{1}$ & $q_{4}$\\
    \midrule
    Bihar & 10.57 & 10.39 && 12.20 & 11.55 && 156.62 & 147.79\\
    Andhra Pradesh & 7.61 & 7.53 && 22.41 & 23.09 && 185.27 & 159.61\\
    Odisha & 12.14 & 10.49 && 16.69 & 21.04 && 163.88 & 118.91 \\
    Madhya Pradesh & 7.38 & 7.16 && 17.70 & 15.63 && 98.82 & 86.31 \\
    Karnataka & 9.76 & 8.75 && 13.52 & 15.87 && 247.25 & 132.07\\
    \bottomrule
  \end{tabular}
    \end{subtable}
\end{table*}

\section{Shapley Dependency Plots}\label{shap_dependency}
\begin{figure*}[hp]
    \centering
    \begin{minipage}{0.19\textwidth}
        \centering
        \includegraphics[width=\textwidth]{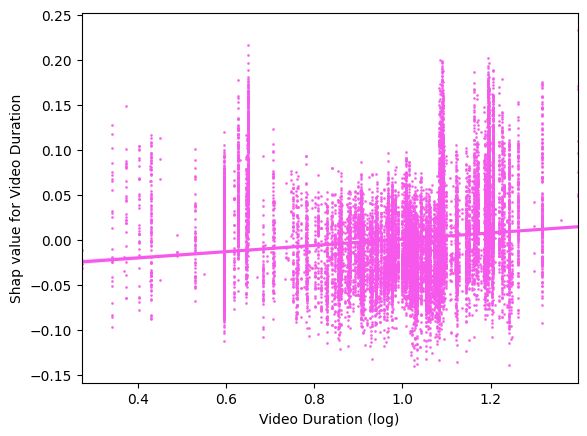}
        \subcaption[first subcaption]{Bihar}\label{fig:video_duration_reg_bihar}
    \end{minipage}
    \begin{minipage}{0.19\textwidth}
        \centering
        \includegraphics[width=\textwidth]{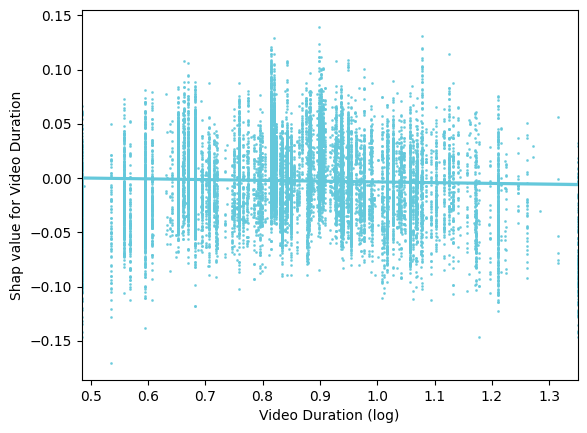}
        \subcaption[second subcaption]{Andhra Pradesh}\label{fig:video_duration_reg_ap}
    \end{minipage}
    \begin{minipage}{0.19\textwidth}
        \centering
        \includegraphics[width=\textwidth]{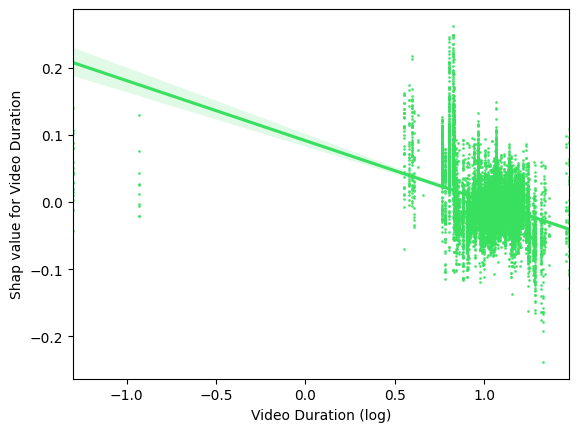}
        \subcaption[second subcaption]{Odisha}\label{fig:video_duration_reg_odisha}
    \end{minipage}
    \begin{minipage}{0.19\textwidth}
        \centering
        \includegraphics[width=\textwidth]{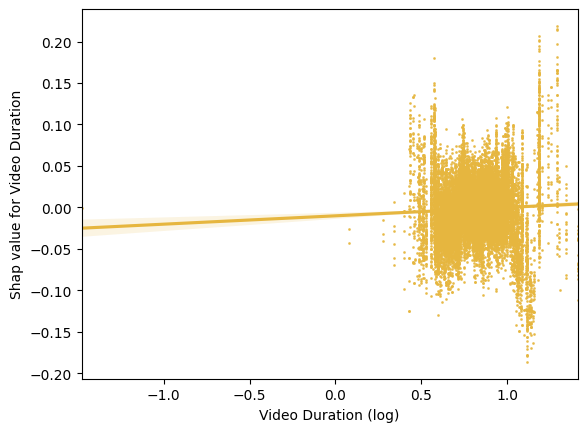}
        \subcaption[second subcaption]{Madhya Pradesh}\label{fig:video_duration_reg_mp}
    \end{minipage}
    \begin{minipage}{0.19\textwidth}
        \centering
        \includegraphics[width=\textwidth]{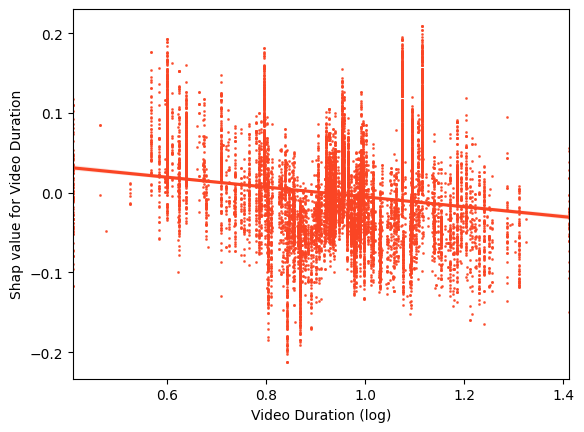}
        \subcaption[second subcaption]{Karnataka}\label{fig:video_duration_reg_karnataka}
    \end{minipage}
    \caption{\textbf{SHAP dependency plots for video duration across the five states when fit with a regression line. Only Bihar and Madhya Pradesh have a positive slope, i.e., videos of longer duration have a positive impact on adoption.}}
    \label{fig:dep_video_duration}
\end{figure*}

\begin{figure*}[hp]
    \centering
    \begin{minipage}{0.19\textwidth}
        \centering
        \includegraphics[width=\textwidth]{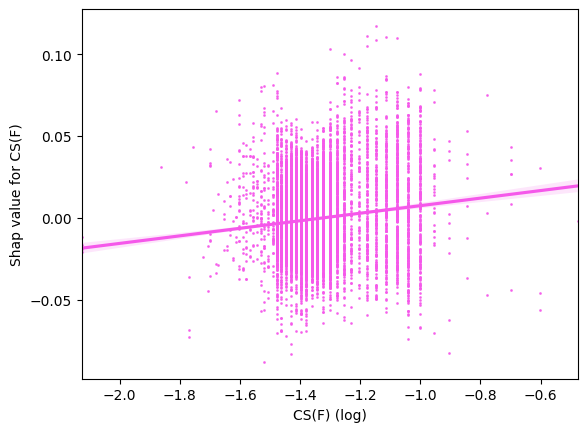}
        \subcaption[first subcaption]{Bihar}\label{fig:csd_bihar}
    \end{minipage}
    \begin{minipage}{0.19\textwidth}
        \centering
        \includegraphics[width=\textwidth]{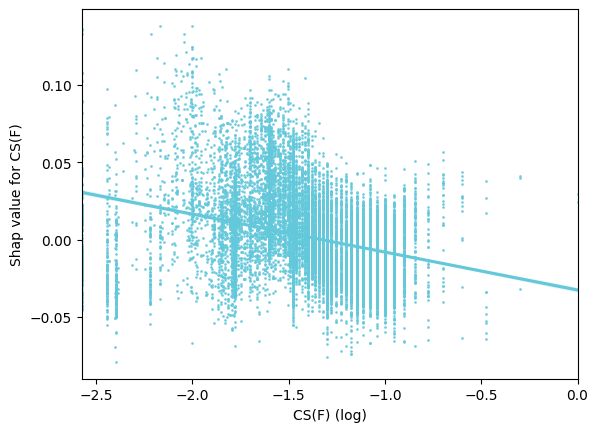}
        \subcaption[second subcaption]{Andhra Pradesh}\label{fig:csd_ap}
    \end{minipage}
    \begin{minipage}{0.19\textwidth}
        \centering
        \includegraphics[width=\textwidth]{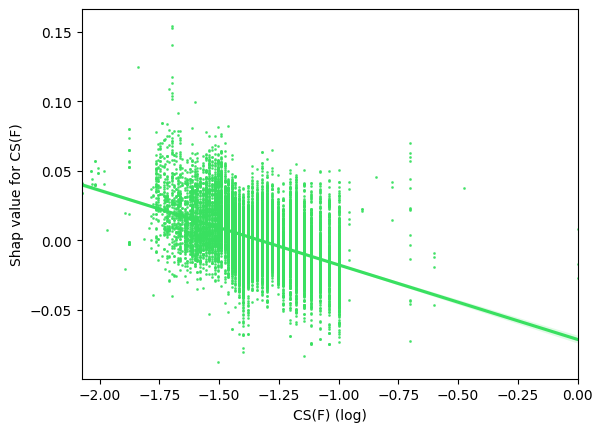}
        \subcaption[second subcaption]{Odisha}\label{fig:csd_odisha}
    \end{minipage}
    \begin{minipage}{0.19\textwidth}
        \centering
        \includegraphics[width=\textwidth]{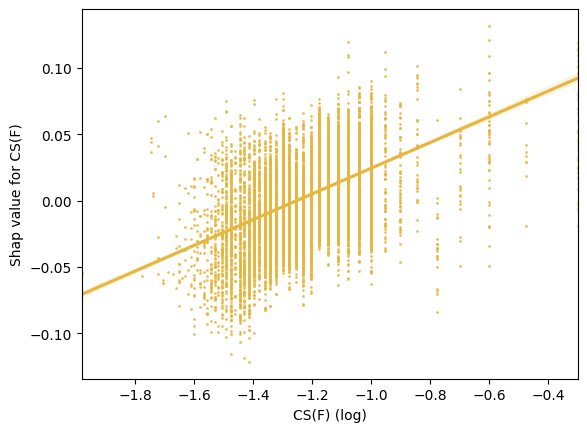}
        \subcaption[second subcaption]{Madhya Pradesh}\label{fig:csd_mp}
    \end{minipage}
    \begin{minipage}{0.19\textwidth}
        \centering
        \includegraphics[width=\textwidth]{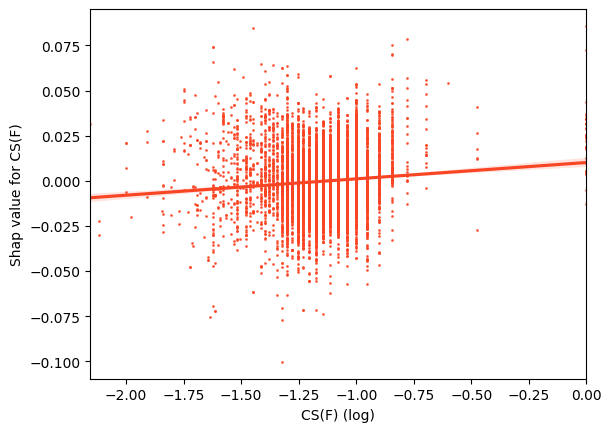}
        \subcaption[second subcaption]{Karnataka}\label{fig:csd_karnataka}
    \end{minipage}
    \caption{\textbf{SHAP dependency plots for $CS_{F}$ across the five states when fit with a regression line. We only see a positive impact on adoptions in Bihar, Madhya Pradesh and Karnataka.}}
    \label{fig:dep_csf}
\end{figure*}

This section contains various SHAP Dependency Plots to capture the relationship between certain features and their impact on the model output in detail. Figure~\ref{fig:dep_video_duration} represents the SHAP dependency plot for video duration. Figures~\ref{fig:dep_csf} represents the SHAP dependency plots for content specificity at the farmer level.

\end{appendices}

\end{document}